\documentclass[amsmath,amssymb,amsfont,prl,groupedaddress,twocolumn,notitlepage, nobibnotes,nofootinbib,longbib]{revtex4-2}
\pdfoutput=1
\usepackage[utf8]{inputenc}
\usepackage[english]{babel}
\usepackage{graphicx}% Include figure files
\usepackage{dcolumn}% Align table columns on decimal point
\usepackage{bm}% bold math
\usepackage{braket}
\usepackage{mathtools}
\usepackage{booktabs}
\usepackage{url}
\usepackage{soul}
\usepackage{array}
\usepackage[colorlinks,citecolor=blue,linkcolor=blue,urlcolor=blue, breaklinks=true]{hyperref}
\usepackage{float}
\usepackage{orcidlink}
\usepackage{bbold}
\usepackage{pdfpages}
\makeatletter
\AtBeginDocument{\let\LS@rot\@undefined}
\makeatother

\newcommand{\dd}{\mathrm{d}\xspace}

\usepackage{xcolor}
\usepackage{xspace}
\usepackage{ifthen}
\graphicspath{{/}}

\newcommand{\showcomments}{true}

\newcommand{\andrea}[1]%
{\ifthenelse{\equal{\showcomments}{true}}%
{{\color{orange}{\small \textbf{Andrea says:} #1}}}{\xspace}}%

\newcommand{\marios}[1]%
{\ifthenelse{\equal{\showcomments}{true}}%
{{\color{blue}{\small \textbf{Marios:} #1}}}{\xspace}}%

\newcommand{\ofek}[1]%
{\ifthenelse{\equal{\showcomments}{true}}%
{{\color{purple}{\small \textbf{Ofek:} #1}}}{\xspace}}%

\newcommand{\markus}[1]%
{\ifthenelse{\equal{\showcomments}{true}}%
{{\color{red}{\small \textbf{Markus:} #1}}}{\xspace}}%

\begin{document}

\begin{abstract}
 Protocols for observing gravity induced entanglement typically comprise the interaction of two particles prepared either in a superposition of two discrete paths, or in a continuously delocalized (harmonic oscillator) state of motion. An important open question has been whether these two different approaches allow to draw the same conclusions on the quantum nature of gravity. To answer this question, we analyse using the path-integral approach a setup that contains both features: a superposition of two highly delocalized center of mass states. We conclude that the two usual protocols are of similar epistemological relevance. In both cases the appearance of entanglement, within linearised quantum gravity, is due to gravity being in a highly non-classical state: a superposition of distinct geometries.
\end{abstract}

\title{Gravity Mediated Entanglement between Oscillators as Quantum Superposition of Geometries}

\author{Ofek Bengyat\,\orcidlink{0000-0001-5547-9176}}
\email{ofek.bengyat@oeaw.ac.at}
% \affiliation{Institute for Quantum Optics and Quantum Information (IQOQI) Vienna, Austrian Academy of Sciences, Boltzmanngasse 3, A-1090 Vienna, Austria}
% \affiliation{Vienna Center for Quantum Science and Technology (VCQ), Faculty of Physics, University of Vienna, Boltzmanngasse 5, A-1090 Vienna, Austria}

\author{Andrea {Di Biagio}\,\orcidlink{0000-0001-9646-8457}}
%\email{andrea.dibiagio@oeaw.ac.at}
% \affiliation{Institute for Quantum Optics and Quantum Information (IQOQI) Vienna, Austrian Academy of Sciences, Boltzmanngasse 3, A-1090 Vienna, Austria}

\author{Markus Aspelmeyer\,\orcidlink{0000-0003-4499-7335}}
% \affiliation{Institute for Quantum Optics and Quantum Information (IQOQI) Vienna, Austrian Academy of Sciences, Boltzmanngasse 3, A-1090 Vienna, Austria}
% \affiliation{Vienna Center for Quantum Science and Technology (VCQ), Faculty of Physics, University of Vienna, Boltzmanngasse 5, A-1090 Vienna, Austria}

\author{Marios Christodoulou\,\orcidlink{0000-0001-6818-2478}}
%\email{marios.christodoulou@univie.ac.at}
\affiliation{Institute for Quantum Optics and Quantum Information (IQOQI) Vienna, Austrian Academy of Sciences, Boltzmanngasse 3, A-1090 Vienna, Austria}
\affiliation{Vienna Center for Quantum Science and Technology (VCQ), Faculty of Physics, University of Vienna, Boltzmanngasse 5, A-1090 Vienna, Austria}

\date{\today}

\maketitle

Feynman argued that detecting the gravitational pull of a mass in superposition would be direct evidence of the quantum nature of gravity \cite{ricklesRoleGravitationPhysics2011}, as the description of such an experiment would require the assignment of quantum mechanical probability amplitudes to different configurations of the gravitational field. Due to the weakness of gravity this has remained a gedankenexperiment since the last 65 years. Partly motivated by experimental progress there has been a revival of this fundamental question and several proposals have been put forward that analyze Feynman's gravity-mediated entanglement (GME) in more detail \cite{boseSpinEntanglementWitness2017,marlettoGravitationallyInducedEntanglement2017,krisnandaObservableQuantumEntanglement2020}. Given the increase in quantum control of mesoscopic solid masses \cite{Poot2012,Aspelmeyer2014,Gonzalez-Ballestero2021} and advances in gravity measurements at small scales \cite{schmoleMicromechanicalProofofprincipleExperiment2016,liuProspectsObservingGravitational2021,westphalMeasurementGravitationalCoupling2021,Lee2020}, these experiments are expected to become feasible in our times \cite{rijavecDecoherenceEffectsNonclassicality2021,aspelmeyerWhenZehMeets2022}.

The literature is split into two kinds of experimental protocols for observing gravity induced entanglement. We refer to them as the \emph{path protocol} and the \emph{oscillator protocol}. The path protocol \cite{boseSpinEntanglementWitness2017,marlettoGravitationallyInducedEntanglement2017} envisages witnessing entanglement production due to the gravitational interaction between two particles, each in a quantum superposition of centre of mass location. The quantum delocalization around each position is neglected. The oscillator protocol \cite{krisnandaObservableQuantumEntanglement2020,yantGravitationalHarmoniumGravitationally2023,boseMechanismQuantumNatured2022} takes a different route: entanglement arises due to gravitational interaction between two continuously delocalized particles, initially prepared as quantum ground states of harmonic oscillators. Both protocols would serve as a test for the predictions of (linearized) quantum gravity.

% \ofek{make this not a footnote.}\footnote{
% Interesting theories in which the gravitational interaction is mediated by a classical field, such as semiclassical gravity (where the curvature of spacetime is sourced by the expectation value of the energy momentum tensor) and Oppenheim's proposal (where the coupling between classical field and quantum matter is stochastic) do not predict entanglement. Linearised quantum gravity predicts entanglement.}

The question remains whether for both protocols one can draw the same conclusion about the quantum nature of gravity. Specifically, that neither case can be described by our existing (classical) theory of gravity, but requires linearized quantum gravity. 

In \cite{christodoulouPossibilityLaboratoryEvidence2019,christodoulouLocallyMediatedEntanglement2023,christodoulouGravityEntanglementQuantum2023,chenQuantumStatesFields2022} it was argued that during the path protocol the entanglement is caused by spacetime existing in a superposition of causally evolving, diffeomorphically inequivalent geometries. In this paper, we extend this analysis to the oscillator protocol. We assume linearised gravity and compute the evolution with the path integral, which has the advantage of keeping spacetime locality explicit. We then consider a state that generalises both the path and oscillator protocols: a superposition of two center of mass states with significant quantum spread of each particle. We calculate the generalised concurrence for this state---an entanglement measure for continuous variable systems proposed in \cite{swainGeneralizedEntanglementMeasure2022}--- in the limit of an instantaneous interaction (zeroth order in $1/c$). We recover results from the literature: in the limit where the width of the wavepackets is small with respect to the size of the path superposition, we recover the entanglement rate for the path protocol. In the converse limit, we recover the entanglement rate for the oscillator protocol. 

We then discuss the interesting case of an initially very localised state, which, perhaps counterintuitively, results in much \emph{faster} entanglement generation. We calculate the leading order relativistic correction for this case, which is a quantitatively larger correction than the relativistic corrections discussed in \cite{christodoulouLocallyMediatedEntanglement2023} for the path protocol, as well as qualitatively different as it arises solely due to the fast spreading of the wavepacket. We conclude that while the physics of the oscillator protocol can in fact be richer, observing gravity mediated entanglement from either the path or oscillator protocol is due to superposition of spacetime geometries.

The computations are analytic but lengthy, involving thousands of terms at times, and so require non-trivial use of a symbolic computation software. The code used is provided as Supplementary Material.

\smallskip

\begin{figure}
    \centering
    \includegraphics[scale=.47]{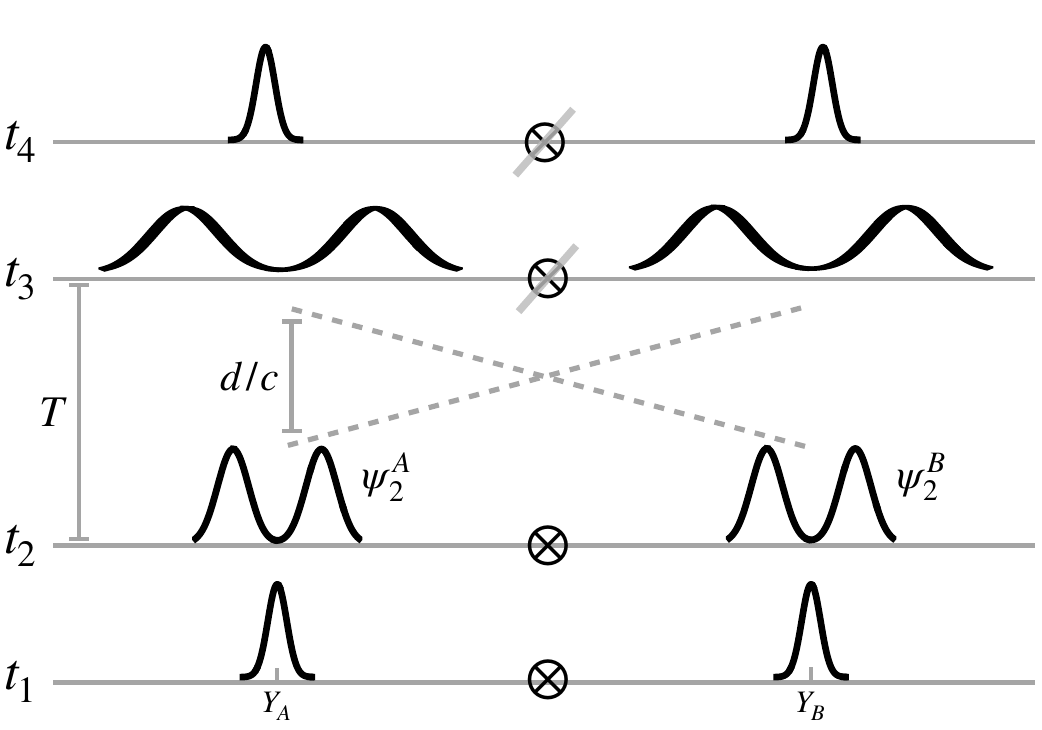}
    \caption{A sketch of the protocol. The particles start in a separable, well-localised state at $t_1$. A more general separable state is prepared by local operations ($t_2$). The particles evolve freely for a period $T$, resulting in a possibly entangled state. Finally, local operations on the particles make them localised again, without changing the entanglement. }
    \label{fig:protocol}
\end{figure}

\paragraph{Protocol ---}
The protocol we study is depicted in Fig.~\ref{fig:protocol}. The initial distance between the center of mass of the particles is denoted $d$. At time $t_1$, each of the two particles $A$ and $B$ is in a well-localized state. This can be achieved by applying a strong trapping potential at times $t<t_1$, which is switched off at time $t_1$. Local operations are applied to each particle to produce some desired wavefunction  $\psi_2(t_2)$, which is assumed separable in $A$ and $B$. The particles are left to interact for a time $T=t_3-t_2$ that is much larger than $t_2-t_1$ and $t_4-t_3$. At time $t_3$, the particles are in some entangled state. Local operations are applied again to the particles to make them well localised (and thus approximately disentangeld from the field) by the time $t_4$. That is, entanglement produced at times $t\notin[t_2,t_3]$ is taken negligible. Entanglement is then measured by measuring center of mass momentum and position correlations.

%%%%%%%%%%%%%%%%%%%%%%%%%%%%%%%%%%%%%%%%%%%%%%%%%%%%%%%%%%%%%%%%
\smallskip
\paragraph{Action ---}
\label{sec:action}
We assume linearized gravity. For brevity, the metric perturbation $h_{\mu\nu}$ at a fixed time is denoted as $\varphi$ and a `path' $h_{\mu\nu}(t)$ as $\mathcal{G}$. The action of the joint system of field and the two particles $A$ and $B$ is of the form
\begin{align}\label{eq:s}
S \left[ x_a , \mathcal{G} \right] = S_\text{G} \left[ x_a , \mathcal{G} \right] + \sum_{a\in\{A,B\}} S_a \left[ x_a \right]
\end{align}
where $a=A,B$, $S_a$ the action of a particle with mass $m_a$ in flat spacetime, and $S_\text{G}$ the gravitational action which includes the kinetic terms for $\varphi$ and its coupling to the energy momentum tensor, which in turn describes point particles with trajectories ${x_a^\mu = (t,\mathbf{x}_a)}$. $S_a$, $S_\text{G}$ and $T_{\mu\nu}$ are given in the Appendix. Note that without the term $S_\text{G}$ there can be no entanglement generation since $e^{i \sum_{a} S_a \left[ x_a \right]/\hbar} = \prod_{a} e^{i  S_a \left[ x_a \right]/\hbar}$.

%%%%%%%%%%%%%%%%%%%%%%%%%%%%%%%%%%%%%%%%%%%%%%%%%%%%%%%%%%%%%%%%%%%%%%%%%%
\smallskip

\emph{Path integral ---} We model the system by a quantum state $ \ket{\Psi(t)} \in \mathcal{H}_A \otimes \mathcal H_B \otimes \mathcal{H}_\text{G}$, where $\mathcal H_A$ and $\mathcal H_B$ are associated with the centre of mass of the particles and $\mathcal{H}_\text{G}$ contains the states of the gravitational field.  The specific assignment of a state to the gravitational field is a delicate point which we discuss further below.

The state at time $t_1$ is assumed to be of the form
\begin{equation}
    \ket{\Psi(t_1)}=\ket{\psi^A_1}\ket{\psi^B_1}\ket{\varphi_1},
\end{equation}
that is, a separable state of the two particles and the field. We also assume that the state at time $t_4$ is
\begin{equation}
    \ket{\Psi(t_4)}= U\ket{\Psi(t_1)}=\ket{\psi_4}\ket{\varphi_1},
\end{equation}
where $U$ is the unitary evolution operator and now $\ket{\psi_4}$ is a possibly non-separable state for the two particle, allowing for gravity mediated entanglement.

Note that we have assumed that the state of the gravitational field is the same at the beginning and at the end of the experiment, for two reasons. First, because the particles are  light and slow-moving, we assume that the amplitude to excite radiation is negligible; in other words, we approximate the final state as its zero graviton component. Secondly, we assume that the particles start and end in well-localised states at the same position. We take $\ket{\varphi_1}$ to be the classical-like state peaked on the classical field $\varphi_1$: the newtonian field\footnote{We take the particles to be localised at the same position $\mathbf Y_a$ at the beginning and end of the experiment. If they would end up in a different place, the final gravitational state would be peaked around a different newtonian field. Nothing but convenience rests on this.} of two localised particles at $\mathbf Y_A$ and $\mathbf Y_B$.

We introduce the path integral by positing that
\begin{equation}
    \bra{\mathbf{y}_4^a,\varphi_1}{U}\ket{\mathbf{y}_1^a,\varphi_1}=\int_{\mathbf{y}_{1}^a,\varphi_1}^{\mathbf{y}_{4}^a,\varphi_1} %\!\!\!\!\!\!\!\!\!\!\!\!\!
    \text{D}x'_a \text{D}\mathcal{G'} e^{i S[x'_a,\mathcal{G'}]},
\end{equation}
where $\ket{\mathbf{y}^a}$ is a position eigenstate for particle $a.$ Therefore the wavefunction for the final state ${\psi_4(\mathbf y_4^a):=\braket{\mathbf{y}^a_4|\psi_4}}$ is given in terms of the initial wavefunctions ${\psi^a_1(\mathbf{y}^a_1):=\braket{\mathbf{y}^a_1|\psi^a_1}}$ as
\begin{align}\label{eq:psi4}
\!\!\!\!\
    \psi_4(\mathbf{y}_{4}^a) = 
    \int \dd \mathbf{y}_{1}^a \psi_1^a(\mathbf{y}_{1}^a)
    \int_{\mathbf{y}_{1}^a,\varphi_1}^{\mathbf{y}_{4}^a,\varphi_1} %\!\!\!\!\!\!\!\!\!\!\!\!\!
    \text{D}x'_a \text{D}\mathcal{G'} e^{i S[x'_a,\mathcal{G'}]}.
\end{align}

% The gravitational field states $\ket{\varphi_1}$ and $\ket{\varphi_4}$ are assumed to be some classical like states peaked on the classical fields $\varphi_1$ and $\varphi_4$. As we have assumed the particles to be well localised at the end and start of the protocol, we take $\varphi_1, \varphi_4$ to be the newtonian fields of two localized particles with mass $m_a$ at locations $\mathbf{y}_{1,a}$ and.\andrea{and?}

% \andrea{add lines here and fix notation}
% The path integral giving the entire evolution reads
% \begin{align}\label{eq:psi4}
% \!\!\!\!\
%     \psi_4(\mathbf{y}_{4}^a) = 
%     \int \dd \mathbf{y}_{1}^a \psi_1^a(\mathbf{y}_{1}^a)
%     \int_{\mathbf{y}_{1}^a,\varphi_1}^{\mathbf{y}_{4}^a,\varphi_1} \!\!\!\!\!\!\!\!\!\!\!\!\!\text{D}x'_a \text{D}\mathcal{G'} e^{i S[x'_a,\mathcal{G'}]}.
% \end{align}

We now take the stationary phase approximation for $\mathcal{G}$. The action is to be evaluated on the classical gravitational field $\mathcal{G}[x_a]$ sourced by a particle moving as $x_a(t)$.  
\begin{align}\label{eq:first_stationary}
    \psi_4(\mathbf{y}_4^a) \approx \int \dd \mathbf{y}_1^a \psi_1(\mathbf{y}_1^a) \int_{\mathbf{y}_1^a}^{\mathbf{y}_4^a} \text{D}x'_a e^{iS[\mathcal{G}[x'_a]]}.
\end{align}
We have introduced the notation ${S[\mathcal{G}[x'_a]]=S[x'_a,\mathcal{G}[x'_a]]}$ for brevity.
Note that, in general, ${\mathcal{G}[x'_a](t_1)\neq\varphi_1\neq\mathcal{G}[x'_a](t_4)}$, therefore the path $\mathcal{G}[x'_a]$ is \emph{not} one of the paths integrated over in \eqref{eq:psi4}. However, we assume for the moment that the error is not too significant, because the initial state will be very localized. The validity of this assumption will be corroborated by our results.  

A second stationary phase approximation for the trajectories is taken by keeping only the contribution on the solution of the classical equations of motion for the particles for initial and final positions $\mathbf{y}_1^a$ and $\mathbf{y}_4^a$, denoted as $x_a$. We obtain that
\begin{align}
\label{eq:generalPoint}
    \psi_4(\mathbf{y}_4^a) \approx \int \dd \mathbf{y}_1^a \psi_1(\mathbf{y}_1^a) e^{iS[\mathcal{G}[x_a]]}.
\end{align}
Since $\psi_1(\mathbf{y}_1^a)$ is separable in $A$ and $B$,  entanglement generation will arise from the $S_G$ part of $S$. The gravitational contribution at each $\mathbf{y}_4^a$ corresponds to a phase ${ S_G[\mathcal{G}[x_a]]}$ with ${x_a(t_1)=\mathbf{y}_1^a}$ and ${x_a(t_4)=\mathbf{y}_4^a}$. Each such pair of trajectories sources a gravitational field, and integrating over $\mathbf{y}_1^a$ corresponds to `summing' the contributions from the gravitational interaction of all these pairs of trajectories for $A$ and $B$.

We note two points that will be further discussed later. First, \eqref{eq:generalPoint} indicates that the gravitational field is not in a classical-like state during the experiment. Second, the trajectories $x_a$ can be taken to be straight lines even when going to the leading order in $1/c$, which indicates no radiation due to no acceleration. This justifies our original assumption of no radiation in $\ket{\varphi_1}$.

%%%%%%%%%%%%%%%%%%%%%%%%%%%%%%%%%%%%%%%%%%%%%%%%%%%%%%%%%%%%%%%%
\smallskip
\paragraph{State evolution ---}
We have yet to use the assumptions that the `splitting' and `recombination' phases are fast with respect to $T$, which imply we only need to consider the free evolution for $t\in [t_2,t_3]$. It can be shown that, to leading order in $1/c$,
\begin{align}
\label{eq:prod}
   \psi_4(\mathbf{y}_4^a) \approx &\int \dd \mathbf{y}_3^a  e^{iS_3[\mathcal{G}[x_3^a]]} \cdot
    \int \dd \mathbf{y}_2^a e^{iS_2[\mathcal{G}[x_2^a]]} 
    \nonumber \\
    &
    \cdot\int \dd \mathbf{y}_1^a \psi_1(\mathbf{y}_1^a) e^{iS_1[\mathcal{G}[x_1^a] ]}.
\end{align}
$x_i^a$ are the trajectories of the particles at the time interval $[t_i,t_{i+1}]$. Above, the actions are approximated up to  leading order in $1/c$ (which is $1/c^2$). We caution that due to retardation it is not straightforward to arrive at \eqref{eq:prod}, see the Appendix. 

Evolution from $\psi_2$ to $\psi_3$ is done by the second integral in \eqref{eq:prod}. The assumptions made previously imply that the evolution done with the actions $S_1$ and $S_3$ will not contribute to the entanglement measured for $\psi_4$. We thus assume that the state $\psi_2 = \int \dd \mathbf{y}_1^a \psi_1 e^{iS_1}$ has been prepared separable in $A$ and $B$ and write 
\begin{equation}
\label{eq:psifi}
   \psi_3(\mathbf{y}_3^a) \approx  \int \dd \mathbf{y}_2^a e^{iS_2[\mathcal{G}[x_2^a]]} \psi_2(\mathbf{y}_2^a).
\end{equation}
Once $\psi_3(\mathbf{y}_3^a)$  is calculated, we proceed to compute the entanglement in $\psi_3$, which in the approximation we work in will be the same as that in $\psi_4$. 
\smallskip

\begin{figure}
    \centering
    \includegraphics[scale=.25]{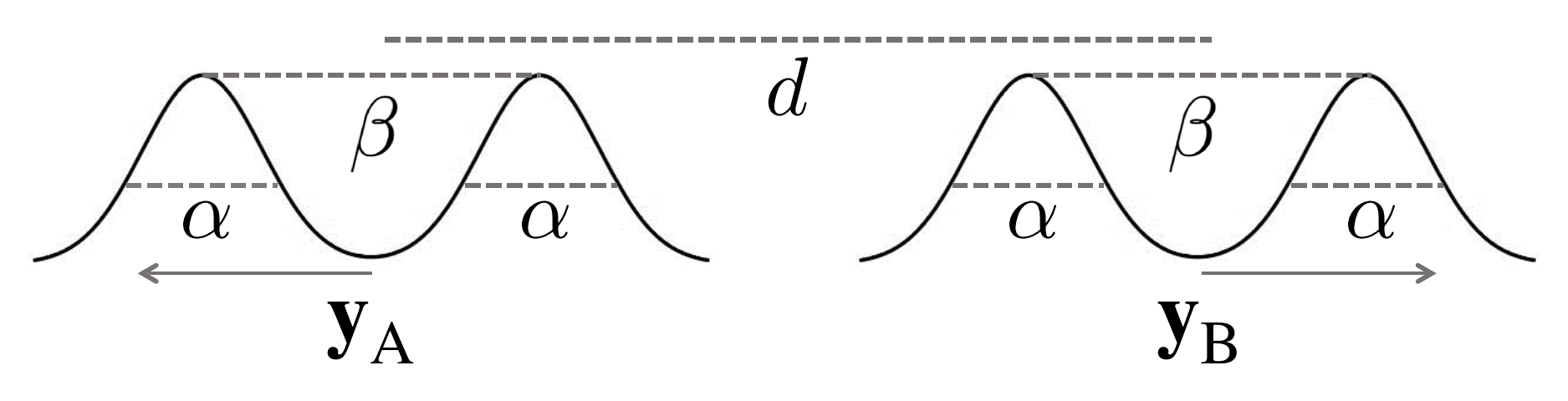}
    \caption{ The initial state of matter. We take the convention that the two particles have opposite positive directions to reduce the number of sign changes in what follows. Note that $\omega\equiv\hbar/m\alpha^2$. }
    \label{fig:twogaussians}
\end{figure}

%%%%%%%%%%%%%%%%%%%%%%%%%%%%%%%%%%%%%%%%%%
\paragraph{Two gaussians in path superposition ---}\label{sec:two_gaussians} We now fix the states $\psi_2^a$ and apply the above to the generalised case shown in Fig.~\ref{fig:twogaussians} to arrive at an analytical expression for the entanglement at zeroth order in $1/c$. That is, in this (and the next) section the gravitational interaction is approximated as an instantaneous interaction through a newtonian potential. The state $\psi_2=\otimes_a \psi_2^a$ is separable in $A$ and $B$ and describes two quantum particles of mass $m_a=m$, each prepared initially in a spatial superposition of two gaussians. The initial wavefunction of each particle in the position basis is
\begin{align}
\label{eq:stateFixed}
    \!\!\!\psi^a_2(\mathbf{y}^a) \propto \exp\left(-\frac{(\mathbf{y}^a - \beta/2)^2}{2\alpha^2}\right) + \exp\left(-\frac{(\mathbf{y}^a + \beta/2)^2}{2\alpha^2}\right),
\end{align}
where $\alpha$ is the delocalization of the position of each path and $\beta$ is the separation between the two paths. The normalisation will drop out from the entanglement.  Thinking of the gaussians as ground states of a harmonic potential (which was turned off at $t=t_2$), we introduce the (nominal) frequency $\omega=\hbar/m\alpha^2$.     

 We can calculate $\psi_3$ using \eqref{eq:psifi}. To simplify the difficult calculation, we assume $d \gg \beta, \alpha, \alpha \omega t$, and expand to second order in the dimensionless parameters $\alpha/d$ and $\beta/d$. This is a realistic assumption for most realistic experimental scenarios. We also expand to zeroth order in the dimensionless parameter $\omega d/c$. The above make the  integrals over $\text{d}\mathbf{y}_2^a$ in \eqref{eq:psifi} computationally tractable. The integrands become sums of gaussians multiplied by polynomials, a case for which closed formulas can be found. The code used for the long analytical calculation is provided as Supplementary Material. 

To quantify the amount of entanglement generated by gravity in $\psi_3$, we compute the generalised entanglement measure $\mathcal{E}$ recently introduced in \cite{swainGeneralizedEntanglementMeasure2022} and given in the Appendix. This is a generalization to the case of continuous sets of states of the concurrence, a well known entanglement measure for discrete sets of states. We find
\begin{align}\label{eq:ent_unified}
    \mathcal{E} =  &\frac{2 G m^2 t}{\hbar d} \frac{1}{d^2} \Biggl[   \frac14\beta^4  + \alpha^2 \beta^2 \left( 1 + \frac14(\omega t)^2 \right) \\\nonumber
        &+ \alpha^4 \biggl[ f_0\left(\frac\beta\alpha\right) +  f_2\left(\frac\beta\alpha\right) \frac{(\omega t)^2}{3} +  f_4\left(\frac\beta\alpha\right) \frac{(\omega t)^4}{9}  \biggr] \Biggr]^{1/2}.
\end{align}
The functions $f_0, f_2$, and $f_4$ are of order unity and given in the appendix. As is shown below, the faster than linear scaling with time is due to the growth in uncertainty.

%%%%%%%%%%%%%%%%%%%%%%%%%%%%%%%%%%%%%
\smallskip
\paragraph{Path and oscillator protocols ---}
\label{sec:path-osc}
When $\beta \gg \alpha$ each of the particles begins in a superposition of two gaussians with negligible overlap. This will remain true during the experiment so long as  $\beta \gg \alpha \omega T $. Then, \eqref{eq:ent_unified} reduces to the path protocol entanglement \cite{aspelmeyerWhenZehMeets2022}
\begin{align}\label{eq:bmv}
    \mathcal{E}_{\beta \gg \alpha} = \frac{G m^2 T}{\hbar d}  \left(  \frac{\beta}{d} \right)^2.
\end{align}
When $\beta \ll \alpha $ each particle state is approximately a single gaussian and \eqref{eq:ent_unified} reduces to 
\begin{align}
\label{eq:oscillator}
    \mathcal{E}_{\beta\ll\alpha}  = \frac{G m^2 T}{\hbar d}  \left(  \frac{\sqrt{2} \alpha}{d} \right)^2 \sqrt{1 + \frac{(\omega T)^2}{3} + \frac{(\omega T)^4}{9}}.
\end{align}
We see that the distance $\beta$ between the paths in the path protocol is replaced by the initial gaussian width $\alpha$. The oscillator protocol has an additional amplification term that depends on $\omega T$. When $\omega T$ is large, we obtain
\begin{align}
\label{eq:oscillatorLargeOmega}
    \mathcal{E}_{\beta\ll\alpha}^{\omega T \gg 1}  = \frac23\frac{G m^2 T}{ \hbar d}  \left(\frac{\alpha\omega T}{d} \right)^2,
\end{align}
which is in accordance with previous literature on the oscillator protocol~\cite{krisnandaObservableQuantumEntanglement2020}.
Recall that $\omega\propto\alpha^{-2}.$ This means that, while in the small $\omega T$ limit, larger $\alpha$ leads to larger entanglement, in the large $\omega T$ limit, smaller spread leads to larger entanglement.
This was already remarked in \cite{krisnandaObservableQuantumEntanglement2020}.
% When $\omega T \gg 1$,  \eqref{eq:oscillator} becomes
% \begin{align}
% \label{eq:oscillatorLargeOmega}
%     \mathcal{E}_{\beta\ll\alpha}^{\omega T \gg 1}  = \frac{G m^2 T}{ \hbar d}  \left(  \frac{\sqrt{2} \alpha(T)}{d} \right)^2,
% \end{align}
% where we have defined $3 \alpha(t) \equiv \alpha \omega t$. 
% Inspecting \eqref{eq:bmv} and \eqref{eq:oscillatorLargeOmega}, entanglement in the path and oscillator protocols is related by $\beta \leftrightarrow \alpha(T)$, where $\alpha(T)$ is the initial spread $\alpha$ magnified by the large number $\omega T$.
% When $\omega T \ll 1$, the spread does not change significantly in time. Accordingly, the entanglement for that case is\ofek{ recovered [instead of ``given"]} given if we replace the path separation $\beta$ with the spread $\sqrt2 \alpha$ in \eqref{eq:bmv}, or, by replacing $\sqrt2 \alpha(T)$ with  $\sqrt2 \alpha$ in \eqref{eq:oscillatorLargeOmega}.\andrea{what?} As already noticed in \cite{krisnandaObservableQuantumEntanglement2020}, in this limit, the entanglement is due to the fast increase with time of $\alpha(t)$. In fact, perhaps counterintuitively, since $3 \alpha(t) = \alpha \omega t=\hbar t/m\alpha$, the smaller the initial spread $\alpha$, the larger will be the entanglement. As we see in the next section, the first correction in $1/c$ for the oscillator protocol captures these physics. 

%%%%%%%%%%%%%%%%%%%%%%%%%%%%%%%%%
\smallskip
\paragraph{Relativistic correction ---}\label{sec:rel} We now report the leading correction in $1/c$ for the oscillator protocol when $\omega T \gg 1$. The newtonian potential is now replaced by a Liénard-Wiechert type \cite{christodoulouLocallyMediatedEntanglement2023}, approximated to leading order in $1/c$. The tedious calculation is given in the Supplementary Material. As in the previous section, we expand the action to second order in $\alpha/d$, but, this time keep the leading order in $\omega d/c$. The state at $t_2$ is now
\begin{align}
    \psi_2 (\mathbf{y}_a) = \otimes_a \psi_2^a(\mathbf{y}_a) \propto \otimes_a \exp\left(-\frac{{\mathbf{y}^a}^2}{2\alpha^2} \right).
\end{align}
 The entangled state $\psi_3$ is computed again through \eqref{eq:psifi} and is given in the Supplementary Material. The entanglement is
 $\mathcal{E}_{\beta\ll\alpha}^{\omega T \gg 1}+ \Delta\mathcal{E}_{\beta\ll\alpha}^{\omega T \gg 1}$ with
\begin{equation}
\label{eq:correction}
     \Delta\mathcal{E}_{\beta\ll\alpha}^{\omega T \gg 1} = -\frac{3}{2} \left( \frac{d}{c t} \right)^2   \mathcal{E}_{\beta\ll\alpha}^{\omega T \gg 1}.  
\end{equation}
Using \eqref{eq:oscillatorLargeOmega}, $\Delta\mathcal{E}_{\beta\ll\alpha}^{\omega t \gg 1}$ scales with $(\omega d/c)^2$. Note that there will be \emph{less} entanglement due to this relativistic effect. The correction for general values of $\omega T$ is given in the Appendix.

There is a physical explanation for this, namely, the fact that when $\omega t \gg 1$ the entanglement generation is not due to the initial spread per se, but, due to the fast growth of the spread with free evolution. The particles see each other with retardation and so they see a smaller spread of the other particle, which results in less entanglement. The amount of entanglement reduction is controlled by the ratio of the distance of the center of mass and the light crossing time.

%\andrea{i have rewritten the following two paragraphs. it still makes the points that (1) the path integral, in this limit, implies continuous superpositions of geometries (2) there is no effect of gravitational attraction. (3) also no radiation. However, i removed the attempt at writing down the state, as it did not make sense to me, or at least it does not follow from the path integral computations}
% \medskip
\smallskip

\paragraph{Gravitational attraction ---} 
Feynman imagined detecting the gravitational field of a mass in superposition by measuring the displacement of another mass. However, surprisingly, gravity mediated entanglement can arise with no measurable displacement. This has been the main insight from the path protocol \cite{boseSpinEntanglementWitness2017,marlettoGravitationallyInducedEntanglement2017}. Owing to our path integral formalism, we can conclude that this is the case for the oscillator protocol too.

To see this, we compare the results of using (i) straight lines trajectories with (ii) the accelerated trajectories of the Kepler problem for the particles. One gets the exact same results at leading order in $1/c$ both for the entanglement \eqref{eq:ent_unified} and all derived results, as well as for the relativistic correction \eqref{eq:correction}.  Therefore at this level of approximation, the entanglement in the oscillator protocol, like that in the path-protocol, is not due to the gravity induced displacement between the particles.

\paragraph{Superposition of geometries ---}
Let us now return to \eqref{eq:generalPoint} and discuss the state of the field. For the scope of this computation, we assume that the state of the field is the same at the beginning and at the end of the experiment. In the evaluation of the path integral, we take both the particles and the fields to be on-shell. The picture that arises, to this approximation, is that of a superposition of point particles on constant-velocity trajectories sourcing a corresponding gravitational field. Different trajectories of the particles will source diffeomorphically inequivalent geometries. Additionally, since the trajectories of the particles can be taken to not be accelerated, these geometries do not contain radiation. Therefore we may say that during the experiment, spacetime is in a superposition of a continuum of different, radiation-free, geometries. We may approximate the state of the particles and field at times $t\in[t_2,t_3]$ as
% We may approximate the state of the particles and field as
% \begin{equation}
%    \ket{\Psi_3} \approx  \int \dd \mathbf{y}_3^a \,\psi_3(\mathbf{y}_3^a)\ket{\mathbf y_3^a}\ket{\varphi(\mathbf{y}_3^a)},
% \end{equation}
% where $\psi_3(\mathbf{y}_3^a)$ is given by \eqref{eq:psifi}, and $\ket{\varphi(\mathbf y_3^a)}$ are semiclassical states of the fields, peaked on classical fields $\varphi(\mathbf{y}_3^a)$ sourced by particles at positions $\mathbf{y}_3^a$.
%\andrea{or
\begin{equation}
   \ket{\Psi(t)} \approx  \int \dd \mathbf{y}^a \,\psi(\mathbf{y}^a)\ket{\mathbf y^a}\ket{\varphi(\mathbf{y}^a)},
\end{equation}
where $\psi(\mathbf y^a)$ is given by \eqref{eq:psifi}, and $\ket{\varphi(\mathbf y^a)}$ are semiclassical states of the fields, peaked on classical fields $\varphi(\mathbf{y}^a)$ sourced by particles at positions $\mathbf{y}^a$.
\smallskip
\paragraph{Conclusions ---} 
We considered a protocol (see Figs. \ref{fig:protocol} and \ref{fig:twogaussians}) that generalises the path and oscillator protocols for gravity mediated entanglement. We calculated the newtonian limit entanglement for this protocol in \eqref{eq:ent_unified} and saw that the path and oscillator protocol entanglement are recovered as physical subregimes in \eqref{eq:bmv} and \eqref{eq:oscillator}. This is the first indication that entanglement in the oscillator and path protocols is due to superposition of spacetime geometries.

 The oscillator and path protocols amount to specific choices for the $\psi^a_2$ wavefunctions for particles $a=A,B$ in \eqref{eq:psifi} (a sum of Dirac deltas for the path protocol and gaussians for the oscillator protocol). Note that \eqref{eq:psifi} and the general setup and procedure used is applicable to more general (continuous or discrete) prepared wavefunctions $\psi^a_2$. We see that the final matter state will be entangled in $A$ and $B$ because different trajectories $x_a$ of the particles source a different classical field and give rise to a different phase in \eqref{eq:psifi}.   

Our computation relies on two approximations typical for GME experiments. One of them is $\alpha/d\ll 1$, meaning that the size of each wavefunction is much smaller than the distance between the two. Another approximation is that $\omega d/c \ll 1$, which means that the experiment's time scales $T, 1/\omega$ are long enough relative to the distance scales $d, \alpha$ such that relativistic effects are negligible. The saddle-point approximation is expected to be valid as long as typical values of the action are much larger than $\hbar$.

Another assumption in our computation is that the initial and final states of the particles are very localised, and that therefore they are disentangled from the field. Of course, even though much harder to identify experimentally, detecting gravitationally induced decoherence due to entanglement with the field \cite{kanno2021noise} would in itself be a great feat. 

% \ofek{caslav wants that we cite Martin-Martinez, i think he means 1507.02688 but i do not see from the abstract how t is related}

Using the path integral formalism, rather than treating the problem as a direct Hamiltonian interaction term as in \cite{krisnandaObservableQuantumEntanglement2020,boseMechanismQuantumNatured2022, aspelmeyerWhenZehMeets2022}, allows us to work with covariant quantities: the actions that feature in the entangling phases in \eqref{eq:generalPoint} and \eqref{eq:psifi}. Thus, in this setting, entanglement is generated in a spacetime-local manner. It follows that if the entire protocol lasts for about time $T$ and $T \ll d/c$ the particles will not become entangled. This was shown by some of us for the path protocol in \cite{christodoulouLocallyMediatedEntanglement2023}. In this work, we applied a similar computational technique to more general wavefunctions.

We reported in \eqref{eq:correction} the first correction in $1/c$ for the oscillator entanglement \eqref{eq:oscillator} for the case $\omega t \gg 1$. In this case the entanglement is generated due to the fast rate of growth of the spread and scales inversely with the initial spread $\alpha$, as noticed in \cite{krisnandaObservableQuantumEntanglement2020}. The relativistic correction we calculated here intuitively corresponds to the particles seeing each other in the past due to retardation and thus interacting with a gaussian with smaller spread. This gives rise to a negative correction with respect to the newtonian calculation, that is, less entanglement. The relative error of the correction versus the newtonian entanglement is $-\frac{3}{2}  \left( d /c T\right)^2$. This relativistic correction is very difficult to detect for gravity, but, it could potentially be observed in an electromagnetic version of the experiment.

Our results for the oscillator protocol without relativistic corrections match those of \cite{krisnandaObservableQuantumEntanglement2020}. Relativistic corrections to the entanglement in the oscillator protocol were also computed in \cite{boseMechanismQuantumNatured2022}, but for a slightly different setup. While we compute the entanglement of an initially separable state as a function of interaction time, they compute the entanglement in the joint ground state of gravity and optically-trapped massive particles. As a result, their entanglement does not depend on time. The time-dependence shown here could be exploited in an experimental configuration that starts in two separable oscillator states and introduces the wanted interaction at a later time\footnote{This also fits better the scope of the discussions of a theory independent argument for the non-classicality of gravity \cite{galleyNogoTheoremNature2022,marlettoWitnessingNonclassicalityQuantum2020, dibiagioRelativisticLocalityCan2023}.}, e.g. by bringing the particles more closely together.

It has been previously argued by some of us in \cite{christodoulouPossibilityLaboratoryEvidence2019,christodoulouLocallyMediatedEntanglement2023, polinoPhotonicImplementationQuantum2022, christodoulouGravityEntanglementQuantum2023} that the path protocol would show that the gravitational field has been indirectly detected in a quantum superposition of macroscopically distinct geometries. An analogous understanding for continuous matter wavefunctions sourcing the gravitational field has thus far been missing. We have shown here  that the entanglement in the oscillator protocol can also be attributed to a superposition of geometries, specifically a superposition of a continuum of different, radiation-free geometries, and the different phases these accrue.

A clear takeaway from our analysis is that if linearised quantum gravity is assumed, whatever the field state is taken to be, it cannot be a classical-like state that approximately solves Einstein's equations.

\section*{acknowledgments}
We acknowledge support of the ID\# 61466 and ID\# 62312 grants from the John Templeton Foundation, as part of the ``Quantum Information Structure of Spacetime (QISS)'' project (\hyperlink{http://www.qiss.fr}{qiss.fr}), and support from the Research Network Quantum Aspects of Spacetime (TURIS). OB acknowledges support from the Blaumann foundation. We thank \v{C}aslav Brukner, Anton Zasedatelev, Sougato Bose, Richard Howl, Anupam Mazumdar and Carlo Rovelli for useful discussions.

%\clearpage
%\onecolumngrid

\section*{Appendix}

\emph{Action ---}
The metric tensor is approximated as ${{g_{\mu\nu} = \eta_{\mu\nu} + h_{\mu\nu}}}$ where $h_{\mu\nu}$ is the metric perturbation that satisfies $|h_{\mu\nu}|\ll 1$ and $\eta_{\mu\nu}$ is the flat metric.
The signature is $(-,+,+,+)$ and we have: $g=\det g_{\mu\nu}$, $t$  the coordinate time, $R$ the Ricci scalar, $G$ Newton's gravitational constant, $c$ the speed of light, and $x_a^\mu = (ct, \mathbf{x}_a(t))$ the trajectory of particle $a$. We have
\begin{equation}
 S_a[x_a]  = - m_a c \int \text{d}t  \sqrt{-\eta_{\mu\nu} \dot{x}_a^\mu \dot{x}_a^\nu} ,
 \end{equation}
\begin{equation}
\begin{aligned}
 S_\text{G} \left[ x_a ,g \right]  =& \int \text{d}^4x \frac{c^3}{16\pi G} [\sqrt{-g} R]_{\mathcal{O}(h_{\mu\nu}^2)} \\
 & - \int \text{d}^4x \frac{1}{2c} h_{\mu\nu} T^{\mu\nu} \big|_{g_{\mu\nu}=\eta_{\mu\nu}} + O(h_{\mu\nu}^3), 
\end{aligned}
\end{equation}
and
\begin{equation}
    T^{\mu\nu} \big|_{g=\eta} = \sum_{a}   \gamma_a(t) m_a  \dot{x}_a^\mu \dot{x}_a^\nu \delta^{(3)}(\mathbf{x}-\mathbf{x}_a).
\end{equation}
Note that taking only the interaction term $\sim T_{\mu\nu} h^{\mu\nu}$ of $S_\text{G}$ would give a wrong numerical result by a factor of two as the contribution of the kinetic terms of $h_{\mu\nu}$ also need to be taken into account. The on-shell action $\tilde{S}_\text{G}$ reads \cite{christodoulouLocallyMediatedEntanglement2023}
\begin{align}\label{eq:action_onshell}
    S_\text{G}[\mathcal{G}[x'_a]] =
    %& \tilde{S}_\text{G} \left(  \mathbf{y}_1 , \mathbf{y}_2 , \mathbf{y}'_1 , \mathbf{y}'_2 \right) = \\\nonumber
    &\sum_{a\neq b} \int_0^t \text{d}t'
     \frac{
     G  m_a  m_b
     \bar{V}_{a}^{\mu\nu}(t_{ab}) V_{b\mu\nu} / c^4   }
    { d_{ab}^\mu \dot{x}'_{a\mu}(t_{ab}) / c},
\end{align}
with
\begin{align}
V_a^{\mu\nu} &= \dot{x}_a^\mu \dot{x}_a^\nu / \sqrt{\dot{x}_a^\mu \dot{x}_a^\nu \eta_{\mu\nu} / c^2},\\
\bar{V}^{\mu\nu} &= V^{\mu\nu} - \frac12 \eta^{\mu\nu} \eta_{\alpha\beta} V^{\alpha\beta},\\
\label{eq:dab}
 d_{ab}^\mu \equiv x_b^\mu - x_a^\mu(&t_{ab}) = (ct-ct_{ab}, \mathbf{x}_b - \mathbf{x}_a(t_{ab})), 
\end{align}
where $t_{ab}$ is the retarded time, defined implicitly as ${d_{ab}^2 =0}$. The omitted time argument in $V_{a\mu\nu}, d_{ab}, \mathbf{x}_a$ and $x_a^\mu$ implies dependence on the non-retarded time $t'$.

\smallskip
\emph{Path integral at leading order in $1/c$ ---}
We split the path integral at times $t_2, t_3$ as
\begin{equation}
\begin{aligned}
    \psi_4(\mathbf{y}_4^a) = &\int \dd \mathbf{y}_3^a\int \dd \mathbf{y}_2^a\int \dd \mathbf{y}_1^a \psi_1(\mathbf{y}_1^a)
      \\
    &\int_{\mathbf{y}_1^a}^{\mathbf{y}_2^a} \text{D}x_1^a e^{iS_1[\mathcal{G}[x_{1}^a]]}
    \int_{\mathbf{y}_2^a}^{\mathbf{y}_3^a} \text{D}x_2^a e^{iS_2[\mathcal{G}[x_1^a,x_2^a]]}   \\
    &\int_{\mathbf{y}_3^a}^{\mathbf{y}_4^a} \text{D}x_3^a e^{iS_3[\mathcal{G}[x_1^a,x_2^a,x_3^a, ]]}.
\end{aligned}
\end{equation}
where the path $x_i$ connects $y_i$ and $y_{i+1}$. %Note the dependence in $S_2$ and $S_3$ on previous paths because of retardation.
Note that, because of the retardation of the field, $S_2$ and $S_3$ depend on the path in the previous time interval. This dependence can be disregarded to leading order in $1/c$, since the time interval influenced by the previous trajectory is then of order $d/c$, which implies that the contribution of the previous trajectories will be one order higher in $1/c$.%\footnote{Another way to see this is that light--tracing back in time for retarded times larger than $d/c$ corresponds to a a spatial part of the wavefunction that is far from the bulk of the probability amplitude. Equivalently, we are taking the `tails' of the gaussians to be approximately zero outside a few standard deviations.}

Schematically, we approximate the action in interval $i$ as
 \begin{align}
     S_i &= \int_{t_i}^{t_{i+1}}  L_{[-\infty,t']}\dd t' \approx \int_{t_i}^{t_{i+1}}  L_{[t'-d/c ,  t']}\dd t'  \\
     &=\int_{t_i}^{t_i+d/c} L_{[t'-d/c ,  t']}\dd t' + \int_{t_i+d/c}^{t_{i+1}} L_{[t'-d/c ,  t']},\dd t' \nonumber
 \end{align}
 where $L_{[t',t'']}$ depends on the particle trajectories between times $t'$ and $t''$. Now, notice that in the first term the integration is over a time interval of length $d/c$. This implies that the contribution from that term will be one order higher in $1/c$ than the second term, meaning we can neglect it. Therefore, after a simple change of variables, we may write
 \begin{align}
     S_i &\approx \int_{t_i}^{t_{i+1}-d/c} L_{[t',  t'+d/c]}.
 \end{align}
Thus $S_i$ only depends on the particle trajectories between $t_i, t_{i+1}$. This way, we can neglect the dependence of $S_2$ and $S_3$ on the past of the trajectories and write $S_3[\mathcal{G}[x_1^a,x_2^a,x_3^a]]\approx S[\mathcal{G}[x_3^a]]$ and $S_2[\mathcal{G}[x_1^a,x_2^a]] \approx S[\mathcal{G}[x_2^a]]$.

\smallskip

\emph{Entanglement measure ---}
The definition of $\mathcal{E}$ for a normalized bi--partite pure state $\ket\psi$ is \cite{swainGeneralizedEntanglementMeasure2022}
\begin{align}
\label{eq:def_ent}
    \mathcal{E}^2 = 2 - 2 \int \text{d}y_1 \text{d}y'_1  \left| \int \text{d}y_2 \psi(y_1,y_2) \psi^*(y'_1,y_2)  \right|^2.
\end{align}

\emph{The functions $f_0$,$f_2$,$f_4$ ---} The functions from Eq.~\eqref{eq:ent_unified} are order unit and given by
\label{app:fs}
% \begin{align}
%     f_0(x) &= \frac{-\frac{x^4}{4}-x^2+e^{\frac{x^2}{2}}-\frac{1}{2} e^{\frac{x^2}{4}} \left(x^4+2 x^2-4\right)+1}{\left(e^{\frac{x^2}{4}}+1\right)^2} \nonumber \\
%     f_2(x) &=\frac{-\frac{3 x^2}{2}+e^{\frac{x^2}{2}}+\left(-\frac{3 x^4}{8}-\frac{3 x^2}{2}+2\right) e^{\frac{x^2}{4}}+1}{\left(e^{\frac{x^2}{4}}+1\right)^2}\\\nonumber
%     f_4(x) &= \frac{\left(-\frac{x^2}{2}+e^{\frac{x^2}{4}}+1\right)^2}{\left(e^{\frac{x^2}{4}}+1\right)^2}.\nonumber
% \end{align}
\begin{align}
    f_0(x) &= \frac{-x^4 - 4x^2 + 4e^{x^2/2} - 2 e^{x^2/4} (x^4 + 2x^2 - 4) + 4}{4(e^{x^2/4} + 1)^2} \\
    f_2(x) &= \frac{-12x^2 + 8e^{x^2/2} + (-3x^4 - 12x^2 + 16) e^{x^2/4} + 8}{8(e^{x^2/4} + 1)^2} \nonumber\\
    f_4(x) &= \frac{(-x^2/2 + e^{x^2/4} + 1)^2}{(e^{x^2/4} + 1)^2} \nonumber
\end{align}

They are plotted in the Supplementary Material.

\emph{Relativistic correction ---} The relativistic correction to \eqref{eq:oscillator} for general values of $\omega T$ was found to be
\begin{align}
     \Delta\mathcal{E}_{\beta\ll\alpha} = & -\frac12 \left( \frac{\omega d}{c} \right)^2  \frac{G m^2 T}{\hbar d} \left( \frac{\sqrt2\alpha}{d} \right)^2  \\\nonumber 
     &\times \frac{\left( \omega T \right)^2 / 3  - 1}{\sqrt{1 + \frac{(\omega T)^2}{3} + \frac{(\omega T)^4}{9}}}.
\end{align}

\bibliography{refs.bib}

\begin{thebibliography}{25}%
\makeatletter
\providecommand \@ifxundefined [1]{%
 \@ifx{#1\undefined}
}%
\providecommand \@ifnum [1]{%
 \ifnum #1\expandafter \@firstoftwo
 \else \expandafter \@secondoftwo
 \fi
}%
\providecommand \@ifx [1]{%
 \ifx #1\expandafter \@firstoftwo
 \else \expandafter \@secondoftwo
 \fi
}%
\providecommand \natexlab [1]{#1}%
\providecommand \enquote  [1]{``#1''}%
\providecommand \bibnamefont  [1]{#1}%
\providecommand \bibfnamefont [1]{#1}%
\providecommand \citenamefont [1]{#1}%
\providecommand \href@noop [0]{\@secondoftwo}%
\providecommand \href [0]{\begingroup \@sanitize@url \@href}%
\providecommand \@href[1]{\@@startlink{#1}\@@href}%
\providecommand \@@href[1]{\endgroup#1\@@endlink}%
\providecommand \@sanitize@url [0]{\catcode `\\12\catcode `\$12\catcode `\&12\catcode `\#12\catcode `\^12\catcode `\_12\catcode `\%12\relax}%
\providecommand \@@startlink[1]{}%
\providecommand \@@endlink[0]{}%
\providecommand \url  [0]{\begingroup\@sanitize@url \@url }%
\providecommand \@url [1]{\endgroup\@href {#1}{\urlprefix }}%
\providecommand \urlprefix  [0]{URL }%
\providecommand \Eprint [0]{\href }%
\providecommand \doibase [0]{https://doi.org/}%
\providecommand \selectlanguage [0]{\@gobble}%
\providecommand \bibinfo  [0]{\@secondoftwo}%
\providecommand \bibfield  [0]{\@secondoftwo}%
\providecommand \translation [1]{[#1]}%
\providecommand \BibitemOpen [0]{}%
\providecommand \bibitemStop [0]{}%
\providecommand \bibitemNoStop [0]{.\EOS\space}%
\providecommand \EOS [0]{\spacefactor3000\relax}%
\providecommand \BibitemShut  [1]{\csname bibitem#1\endcsname}%
\let\auto@bib@innerbib\@empty
%</preamble>
\bibitem [{\citenamefont {Rickles}\ and\ \citenamefont {DeWitt}(2011)}]{ricklesRoleGravitationPhysics2011}%
  \BibitemOpen
  \bibinfo {editor} {\bibfnamefont {D.}~\bibnamefont {Rickles}}\ and\ \bibinfo {editor} {\bibfnamefont {C.~M.}\ \bibnamefont {DeWitt}},\ eds.,\ \href {https://doi.org/10.34663/9783945561294-00} {\emph {\bibinfo {title} {The {{Role}} of {{Gravitation}} in {{Physics}}: {{Report}} from the 1957 {{Chapel Hill Conference}}}}},\ {{EOS}} \textendash{} {{Sources}}\ (\bibinfo  {publisher} {{Max-Planck-Gesellschaft zur F\"orderung der Wissenschaften}},\ \bibinfo {address} {{Berlin}},\ \bibinfo {year} {2011})\BibitemShut {NoStop}%
\bibitem [{\citenamefont {Bose}\ \emph {et~al.}(2017)\citenamefont {Bose}, \citenamefont {Mazumdar}, \citenamefont {Morley}, \citenamefont {Ulbricht}, \citenamefont {Toro{\v s}}, \citenamefont {Paternostro}, \citenamefont {Geraci}, \citenamefont {Barker}, \citenamefont {Kim},\ and\ \citenamefont {Milburn}}]{boseSpinEntanglementWitness2017}%
  \BibitemOpen
  \bibfield  {author} {\bibinfo {author} {\bibfnamefont {S.}~\bibnamefont {Bose}}, \bibinfo {author} {\bibfnamefont {A.}~\bibnamefont {Mazumdar}}, \bibinfo {author} {\bibfnamefont {G.~W.}\ \bibnamefont {Morley}}, \bibinfo {author} {\bibfnamefont {H.}~\bibnamefont {Ulbricht}}, \bibinfo {author} {\bibfnamefont {M.}~\bibnamefont {Toro{\v s}}}, \bibinfo {author} {\bibfnamefont {M.}~\bibnamefont {Paternostro}}, \bibinfo {author} {\bibfnamefont {A.~A.}\ \bibnamefont {Geraci}}, \bibinfo {author} {\bibfnamefont {P.~F.}\ \bibnamefont {Barker}}, \bibinfo {author} {\bibfnamefont {M.~S.}\ \bibnamefont {Kim}},\ and\ \bibinfo {author} {\bibfnamefont {G.}~\bibnamefont {Milburn}},\ }\bibfield  {title} {\bibinfo {title} {Spin {{Entanglement Witness}} for {{Quantum Gravity}}},\ }\href {https://doi.org/10.1103/PhysRevLett.119.240401} {\bibfield  {journal} {\bibinfo  {journal} {Physical Review Letters}\ }\textbf {\bibinfo {volume} {119}},\ \bibinfo {pages} {240401} (\bibinfo {year} {2017})}\BibitemShut {NoStop}%
\bibitem [{\citenamefont {Marletto}\ and\ \citenamefont {Vedral}(2017)}]{marlettoGravitationallyInducedEntanglement2017}%
  \BibitemOpen
  \bibfield  {author} {\bibinfo {author} {\bibfnamefont {C.}~\bibnamefont {Marletto}}\ and\ \bibinfo {author} {\bibfnamefont {V.}~\bibnamefont {Vedral}},\ }\bibfield  {title} {\bibinfo {title} {Gravitationally {{Induced Entanglement}} between {{Two Massive Particles}} is {{Sufficient Evidence}} of {{Quantum Effects}} in {{Gravity}}},\ }\href {https://doi.org/10.1103/PhysRevLett.119.240402} {\bibfield  {journal} {\bibinfo  {journal} {Physical Review Letters}\ }\textbf {\bibinfo {volume} {119}},\ \bibinfo {pages} {240402} (\bibinfo {year} {2017})}\BibitemShut {NoStop}%
\bibitem [{\citenamefont {Krisnanda}\ \emph {et~al.}(2020)\citenamefont {Krisnanda}, \citenamefont {Tham}, \citenamefont {Paternostro},\ and\ \citenamefont {Paterek}}]{krisnandaObservableQuantumEntanglement2020}%
  \BibitemOpen
  \bibfield  {author} {\bibinfo {author} {\bibfnamefont {T.}~\bibnamefont {Krisnanda}}, \bibinfo {author} {\bibfnamefont {G.~Y.}\ \bibnamefont {Tham}}, \bibinfo {author} {\bibfnamefont {M.}~\bibnamefont {Paternostro}},\ and\ \bibinfo {author} {\bibfnamefont {T.}~\bibnamefont {Paterek}},\ }\bibfield  {title} {\bibinfo {title} {Observable quantum entanglement due to gravity},\ }\href {https://doi.org/10.1038/s41534-020-0243-y} {\bibfield  {journal} {\bibinfo  {journal} {npj Quantum Information}\ }\textbf {\bibinfo {volume} {6}},\ \bibinfo {pages} {1} (\bibinfo {year} {2020})}\BibitemShut {NoStop}%
\bibitem [{\citenamefont {Poot}\ and\ \citenamefont {van~der Zant}(2012)}]{Poot2012}%
  \BibitemOpen
  \bibfield  {author} {\bibinfo {author} {\bibfnamefont {M.}~\bibnamefont {Poot}}\ and\ \bibinfo {author} {\bibfnamefont {H.~S.}\ \bibnamefont {van~der Zant}},\ }\bibfield  {title} {\bibinfo {title} {{Mechanical systems in the quantum regime}},\ }\href {https://doi.org/10.1016/j.physrep.2011.12.004} {\bibfield  {journal} {\bibinfo  {journal} {Physics Reports}\ }\textbf {\bibinfo {volume} {511}},\ \bibinfo {pages} {273} (\bibinfo {year} {2012})}\BibitemShut {NoStop}%
\bibitem [{\citenamefont {Aspelmeyer}\ \emph {et~al.}(2014)\citenamefont {Aspelmeyer}, \citenamefont {Kippenberg},\ and\ \citenamefont {Marquardt}}]{Aspelmeyer2014}%
  \BibitemOpen
  \bibfield  {author} {\bibinfo {author} {\bibfnamefont {M.}~\bibnamefont {Aspelmeyer}}, \bibinfo {author} {\bibfnamefont {T.~J.}\ \bibnamefont {Kippenberg}},\ and\ \bibinfo {author} {\bibfnamefont {F.}~\bibnamefont {Marquardt}},\ }\bibfield  {title} {\bibinfo {title} {{Cavity optomechanics}},\ }\href {https://doi.org/http://dx.doi.org/10.1103/RevModPhys.86.1391} {\bibfield  {journal} {\bibinfo  {journal} {Rev.\ Mod.\ Phys.}\ }\textbf {\bibinfo {volume} {86}},\ \bibinfo {pages} {1391} (\bibinfo {year} {2014})}\BibitemShut {NoStop}%
\bibitem [{\citenamefont {Gonzalez-Ballestero}\ \emph {et~al.}(2021)\citenamefont {Gonzalez-Ballestero}, \citenamefont {Aspelmeyer}, \citenamefont {Novotny}, \citenamefont {Quidant},\ and\ \citenamefont {Romero-Isart}}]{Gonzalez-Ballestero2021}%
  \BibitemOpen
  \bibfield  {author} {\bibinfo {author} {\bibfnamefont {C.}~\bibnamefont {Gonzalez-Ballestero}}, \bibinfo {author} {\bibfnamefont {M.}~\bibnamefont {Aspelmeyer}}, \bibinfo {author} {\bibfnamefont {L.}~\bibnamefont {Novotny}}, \bibinfo {author} {\bibfnamefont {R.}~\bibnamefont {Quidant}},\ and\ \bibinfo {author} {\bibfnamefont {O.}~\bibnamefont {Romero-Isart}},\ }\bibfield  {title} {\bibinfo {title} {{Levitodynamics: Levitation and control of microscopic objects in vacuum}},\ }\bibfield  {journal} {\bibinfo  {journal} {Science}\ }\textbf {\bibinfo {volume} {374}},\ \href {https://doi.org/10.1126/science.abg3027} {10.1126/science.abg3027} (\bibinfo {year} {2021}),\ \Eprint {https://arxiv.org/abs/2111.05215} {arXiv:2111.05215} \BibitemShut {NoStop}%
\bibitem [{\citenamefont {Schm{\"o}le}\ \emph {et~al.}(2016)\citenamefont {Schm{\"o}le}, \citenamefont {Dragosits}, \citenamefont {Hepach},\ and\ \citenamefont {Aspelmeyer}}]{schmoleMicromechanicalProofofprincipleExperiment2016}%
  \BibitemOpen
  \bibfield  {author} {\bibinfo {author} {\bibfnamefont {J.}~\bibnamefont {Schm{\"o}le}}, \bibinfo {author} {\bibfnamefont {M.}~\bibnamefont {Dragosits}}, \bibinfo {author} {\bibfnamefont {H.}~\bibnamefont {Hepach}},\ and\ \bibinfo {author} {\bibfnamefont {M.}~\bibnamefont {Aspelmeyer}},\ }\bibfield  {title} {\bibinfo {title} {A micromechanical proof-of-principle experiment for measuring the gravitational force of milligram masses},\ }\href {https://doi.org/10.1088/0264-9381/33/12/125031} {\bibfield  {journal} {\bibinfo  {journal} {Classical and Quantum Gravity}\ }\textbf {\bibinfo {volume} {33}},\ \bibinfo {pages} {125031} (\bibinfo {year} {2016})},\ \Eprint {https://arxiv.org/abs/1602.07539} {arXiv:1602.07539} \BibitemShut {NoStop}%
\bibitem [{\citenamefont {Liu}\ \emph {et~al.}(2021)\citenamefont {Liu}, \citenamefont {Mummery},\ and\ \citenamefont {Sillanp{\"a}{\"a}}}]{liuProspectsObservingGravitational2021}%
  \BibitemOpen
  \bibfield  {author} {\bibinfo {author} {\bibfnamefont {Y.}~\bibnamefont {Liu}}, \bibinfo {author} {\bibfnamefont {J.}~\bibnamefont {Mummery}},\ and\ \bibinfo {author} {\bibfnamefont {M.~A.}\ \bibnamefont {Sillanp{\"a}{\"a}}},\ }\bibfield  {title} {\bibinfo {title} {Prospects for observing gravitational forces between nonclassical mechanical oscillators},\ }\href {https://doi.org/10.1103/PhysRevApplied.15.034004} {\bibfield  {journal} {\bibinfo  {journal} {Physical Review Applied}\ }\textbf {\bibinfo {volume} {15}},\ \bibinfo {pages} {034004} (\bibinfo {year} {2021})},\ \Eprint {https://arxiv.org/abs/2008.10477} {arXiv:2008.10477} \BibitemShut {NoStop}%
\bibitem [{\citenamefont {Westphal}\ \emph {et~al.}(2021)\citenamefont {Westphal}, \citenamefont {Hepach}, \citenamefont {Pfaff},\ and\ \citenamefont {Aspelmeyer}}]{westphalMeasurementGravitationalCoupling2021}%
  \BibitemOpen
  \bibfield  {author} {\bibinfo {author} {\bibfnamefont {T.}~\bibnamefont {Westphal}}, \bibinfo {author} {\bibfnamefont {H.}~\bibnamefont {Hepach}}, \bibinfo {author} {\bibfnamefont {J.}~\bibnamefont {Pfaff}},\ and\ \bibinfo {author} {\bibfnamefont {M.}~\bibnamefont {Aspelmeyer}},\ }\bibfield  {title} {\bibinfo {title} {Measurement of gravitational coupling between millimetre-sized masses},\ }\href {https://doi.org/10.1038/s41586-021-03250-7} {\bibfield  {journal} {\bibinfo  {journal} {Nature}\ }\textbf {\bibinfo {volume} {591}},\ \bibinfo {pages} {225} (\bibinfo {year} {2021})}\BibitemShut {NoStop}%
\bibitem [{\citenamefont {Lee}\ \emph {et~al.}(2020)\citenamefont {Lee}, \citenamefont {Adelberger}, \citenamefont {Cook}, \citenamefont {Fleischer},\ and\ \citenamefont {Heckel}}]{Lee2020}%
  \BibitemOpen
  \bibfield  {author} {\bibinfo {author} {\bibfnamefont {J.~G.}\ \bibnamefont {Lee}}, \bibinfo {author} {\bibfnamefont {E.~G.}\ \bibnamefont {Adelberger}}, \bibinfo {author} {\bibfnamefont {T.~S.}\ \bibnamefont {Cook}}, \bibinfo {author} {\bibfnamefont {S.~M.}\ \bibnamefont {Fleischer}},\ and\ \bibinfo {author} {\bibfnamefont {B.~R.}\ \bibnamefont {Heckel}},\ }\bibfield  {title} {\bibinfo {title} {{New Test of the Gravitational $1/r^2$ Law at Separations down to 52 $\mu$m}},\ }\href {https://doi.org/10.1103/PhysRevLett.124.101101} {\bibfield  {journal} {\bibinfo  {journal} {Physical Review Letters}\ }\textbf {\bibinfo {volume} {124}},\ \bibinfo {pages} {101101} (\bibinfo {year} {2020})}\BibitemShut {NoStop}%
\bibitem [{\citenamefont {Rijavec}\ \emph {et~al.}(2021)\citenamefont {Rijavec}, \citenamefont {Carlesso}, \citenamefont {Bassi}, \citenamefont {Vedral},\ and\ \citenamefont {Marletto}}]{rijavecDecoherenceEffectsNonclassicality2021}%
  \BibitemOpen
  \bibfield  {author} {\bibinfo {author} {\bibfnamefont {S.}~\bibnamefont {Rijavec}}, \bibinfo {author} {\bibfnamefont {M.}~\bibnamefont {Carlesso}}, \bibinfo {author} {\bibfnamefont {A.}~\bibnamefont {Bassi}}, \bibinfo {author} {\bibfnamefont {V.}~\bibnamefont {Vedral}},\ and\ \bibinfo {author} {\bibfnamefont {C.}~\bibnamefont {Marletto}},\ }\bibfield  {title} {\bibinfo {title} {Decoherence effects in non-classicality tests of gravity},\ }\href {https://doi.org/10.1088/1367-2630/abf3eb} {\bibfield  {journal} {\bibinfo  {journal} {New Journal of Physics}\ }\textbf {\bibinfo {volume} {23}},\ \bibinfo {pages} {043040} (\bibinfo {year} {2021})},\ \Eprint {https://arxiv.org/abs/2012.06230} {arXiv:2012.06230} \BibitemShut {NoStop}%
\bibitem [{\citenamefont {Aspelmeyer}(2022)}]{aspelmeyerWhenZehMeets2022}%
  \BibitemOpen
  \bibfield  {author} {\bibinfo {author} {\bibfnamefont {M.}~\bibnamefont {Aspelmeyer}},\ }\bibfield  {title} {\bibinfo {title} {When {{Zeh Meets Feynman}}: {{How}} to {{Avoid}} the {{Appearance}} of a {{Classical World}} in {{Gravity Experiments}}},\ }in\ \href {https://doi.org/10.1007/978-3-030-88781-0_5} {\emph {\bibinfo {booktitle} {From {{Quantum}} to {{Classical}}}}},\ Vol.\ \bibinfo {volume} {204},\ \bibinfo {editor} {edited by\ \bibinfo {editor} {\bibfnamefont {C.}~\bibnamefont {Kiefer}}}\ (\bibinfo  {publisher} {{Springer International Publishing}},\ \bibinfo {address} {{Cham}},\ \bibinfo {year} {2022})\ pp.\ \bibinfo {pages} {85--95}\BibitemShut {NoStop}%
\bibitem [{\citenamefont {Yant}\ and\ \citenamefont {Blencowe}(2023)}]{yantGravitationalHarmoniumGravitationally2023}%
  \BibitemOpen
  \bibfield  {author} {\bibinfo {author} {\bibfnamefont {J.}~\bibnamefont {Yant}}\ and\ \bibinfo {author} {\bibfnamefont {M.}~\bibnamefont {Blencowe}},\ }\href@noop {} {\bibinfo {title} {Gravitational {{Harmonium}}: {{Gravitationally Induced Entanglement}} in a {{Harmonic Trap}}}} (\bibinfo {year} {2023}),\ \Eprint {https://arxiv.org/abs/2302.05463} {arXiv:2302.05463} \BibitemShut {NoStop}%
\bibitem [{\citenamefont {Bose}\ \emph {et~al.}(2022)\citenamefont {Bose}, \citenamefont {Mazumdar}, \citenamefont {Schut},\ and\ \citenamefont {Toro{\v s}}}]{boseMechanismQuantumNatured2022}%
  \BibitemOpen
  \bibfield  {author} {\bibinfo {author} {\bibfnamefont {S.}~\bibnamefont {Bose}}, \bibinfo {author} {\bibfnamefont {A.}~\bibnamefont {Mazumdar}}, \bibinfo {author} {\bibfnamefont {M.}~\bibnamefont {Schut}},\ and\ \bibinfo {author} {\bibfnamefont {M.}~\bibnamefont {Toro{\v s}}},\ }\bibfield  {title} {\bibinfo {title} {Mechanism for the quantum natured gravitons to entangle masses},\ }\href {https://doi.org/10.1103/PhysRevD.105.106028} {\bibfield  {journal} {\bibinfo  {journal} {Physical Review D}\ }\textbf {\bibinfo {volume} {105}},\ \bibinfo {pages} {106028} (\bibinfo {year} {2022})}\BibitemShut {NoStop}%
\bibitem [{\citenamefont {Christodoulou}\ and\ \citenamefont {Rovelli}(2019)}]{christodoulouPossibilityLaboratoryEvidence2019}%
  \BibitemOpen
  \bibfield  {author} {\bibinfo {author} {\bibfnamefont {M.}~\bibnamefont {Christodoulou}}\ and\ \bibinfo {author} {\bibfnamefont {C.}~\bibnamefont {Rovelli}},\ }\bibfield  {title} {\bibinfo {title} {On the possibility of laboratory evidence for quantum superposition of geometries},\ }\href {https://doi.org/10.1016/j.physletb.2019.03.015} {\bibfield  {journal} {\bibinfo  {journal} {Physics Letters B}\ }\textbf {\bibinfo {volume} {792}},\ \bibinfo {pages} {64} (\bibinfo {year} {2019})},\ \Eprint {https://arxiv.org/abs/1808.05842} {arXiv:1808.05842} \BibitemShut {NoStop}%
\bibitem [{\citenamefont {Christodoulou}\ \emph {et~al.}(2023{\natexlab{a}})\citenamefont {Christodoulou}, \citenamefont {Di~Biagio}, \citenamefont {Aspelmeyer}, \citenamefont {Brukner}, \citenamefont {Rovelli},\ and\ \citenamefont {Howl}}]{christodoulouLocallyMediatedEntanglement2023}%
  \BibitemOpen
  \bibfield  {author} {\bibinfo {author} {\bibfnamefont {M.}~\bibnamefont {Christodoulou}}, \bibinfo {author} {\bibfnamefont {A.}~\bibnamefont {Di~Biagio}}, \bibinfo {author} {\bibfnamefont {M.}~\bibnamefont {Aspelmeyer}}, \bibinfo {author} {\bibfnamefont {{\v C}.}~\bibnamefont {Brukner}}, \bibinfo {author} {\bibfnamefont {C.}~\bibnamefont {Rovelli}},\ and\ \bibinfo {author} {\bibfnamefont {R.}~\bibnamefont {Howl}},\ }\bibfield  {title} {\bibinfo {title} {Locally {{Mediated Entanglement}} in {{Linearized Quantum Gravity}}},\ }\href {https://doi.org/10.1103/PhysRevLett.130.100202} {\bibfield  {journal} {\bibinfo  {journal} {Physical Review Letters}\ }\textbf {\bibinfo {volume} {130}},\ \bibinfo {pages} {100202} (\bibinfo {year} {2023}{\natexlab{a}})}\BibitemShut {NoStop}%
\bibitem [{\citenamefont {Christodoulou}\ \emph {et~al.}(2023{\natexlab{b}})\citenamefont {Christodoulou}, \citenamefont {Biagio}, \citenamefont {Howl},\ and\ \citenamefont {Rovelli}}]{christodoulouGravityEntanglementQuantum2023}%
  \BibitemOpen
  \bibfield  {author} {\bibinfo {author} {\bibfnamefont {M.}~\bibnamefont {Christodoulou}}, \bibinfo {author} {\bibfnamefont {A.~D.}\ \bibnamefont {Biagio}}, \bibinfo {author} {\bibfnamefont {R.}~\bibnamefont {Howl}},\ and\ \bibinfo {author} {\bibfnamefont {C.}~\bibnamefont {Rovelli}},\ }\bibfield  {title} {\bibinfo {title} {Gravity entanglement, quantum reference systems, degrees of freedom},\ }\href {https://doi.org/10.1088/1361-6382/acb0aa} {\bibfield  {journal} {\bibinfo  {journal} {Classical and Quantum Gravity}\ }\textbf {\bibinfo {volume} {40}},\ \bibinfo {pages} {047001} (\bibinfo {year} {2023}{\natexlab{b}})}\BibitemShut {NoStop}%
\bibitem [{\citenamefont {Chen}\ \emph {et~al.}(2022)\citenamefont {Chen}, \citenamefont {Giacomini},\ and\ \citenamefont {Rovelli}}]{chenQuantumStatesFields2022}%
  \BibitemOpen
  \bibfield  {author} {\bibinfo {author} {\bibfnamefont {L.-Q.}\ \bibnamefont {Chen}}, \bibinfo {author} {\bibfnamefont {F.}~\bibnamefont {Giacomini}},\ and\ \bibinfo {author} {\bibfnamefont {C.}~\bibnamefont {Rovelli}},\ }\href@noop {} {\bibinfo {title} {Quantum {{States}} of {{Fields}} for {{Quantum Split Sources}}}} (\bibinfo {year} {2022}),\ \Eprint {https://arxiv.org/abs/2207.10592} {arXiv:2207.10592} \BibitemShut {NoStop}%
\bibitem [{\citenamefont {Swain}\ \emph {et~al.}(2022)\citenamefont {Swain}, \citenamefont {Bhaskara},\ and\ \citenamefont {Panigrahi}}]{swainGeneralizedEntanglementMeasure2022}%
  \BibitemOpen
  \bibfield  {author} {\bibinfo {author} {\bibfnamefont {S.~N.}\ \bibnamefont {Swain}}, \bibinfo {author} {\bibfnamefont {V.~S.}\ \bibnamefont {Bhaskara}},\ and\ \bibinfo {author} {\bibfnamefont {P.~K.}\ \bibnamefont {Panigrahi}},\ }\bibfield  {title} {\bibinfo {title} {Generalized {{Entanglement Measure}} for {{Continuous Variable Systems}}},\ }\href {https://doi.org/10.1103/PhysRevA.105.052441} {\bibfield  {journal} {\bibinfo  {journal} {Physical Review A}\ }\textbf {\bibinfo {volume} {105}},\ \bibinfo {pages} {052441} (\bibinfo {year} {2022})},\ \Eprint {https://arxiv.org/abs/1706.01448} {arXiv:1706.01448} \BibitemShut {NoStop}%
\bibitem [{\citenamefont {Kanno}\ \emph {et~al.}(2021)\citenamefont {Kanno}, \citenamefont {Soda},\ and\ \citenamefont {Tokuda}}]{kanno2021noise}%
  \BibitemOpen
  \bibfield  {author} {\bibinfo {author} {\bibfnamefont {S.}~\bibnamefont {Kanno}}, \bibinfo {author} {\bibfnamefont {J.}~\bibnamefont {Soda}},\ and\ \bibinfo {author} {\bibfnamefont {J.}~\bibnamefont {Tokuda}},\ }\bibfield  {title} {\bibinfo {title} {Noise and decoherence induced by gravitons},\ }\href {https://doi.org/10.1103/PhysRevD.103.044017} {\bibfield  {journal} {\bibinfo  {journal} {Physical Review D}\ }\textbf {\bibinfo {volume} {103}},\ \bibinfo {pages} {044017} (\bibinfo {year} {2021})},\ \Eprint {https://arxiv.org/abs/2007.09838} {arXiv:2007.09838} \BibitemShut {NoStop}%
\bibitem [{\citenamefont {Galley}\ \emph {et~al.}(2022)\citenamefont {Galley}, \citenamefont {Giacomini},\ and\ \citenamefont {Selby}}]{galleyNogoTheoremNature2022}%
  \BibitemOpen
  \bibfield  {author} {\bibinfo {author} {\bibfnamefont {T.~D.}\ \bibnamefont {Galley}}, \bibinfo {author} {\bibfnamefont {F.}~\bibnamefont {Giacomini}},\ and\ \bibinfo {author} {\bibfnamefont {J.~H.}\ \bibnamefont {Selby}},\ }\bibfield  {title} {\bibinfo {title} {A no-go theorem on the nature of the gravitational field beyond quantum theory},\ }\href {https://doi.org/10.22331/q-2022-08-17-779} {\bibfield  {journal} {\bibinfo  {journal} {Quantum}\ }\textbf {\bibinfo {volume} {6}},\ \bibinfo {pages} {779} (\bibinfo {year} {2022})}\BibitemShut {NoStop}%
\bibitem [{\citenamefont {Marletto}\ and\ \citenamefont {Vedral}(2020)}]{marlettoWitnessingNonclassicalityQuantum2020}%
  \BibitemOpen
  \bibfield  {author} {\bibinfo {author} {\bibfnamefont {C.}~\bibnamefont {Marletto}}\ and\ \bibinfo {author} {\bibfnamefont {V.}~\bibnamefont {Vedral}},\ }\bibfield  {title} {\bibinfo {title} {Witnessing nonclassicality beyond quantum theory},\ }\href {https://doi.org/10.1103/PhysRevD.102.086012} {\bibfield  {journal} {\bibinfo  {journal} {Physical Review D: Particles and Fields}\ }\textbf {\bibinfo {volume} {102}},\ \bibinfo {pages} {086012} (\bibinfo {year} {2020})},\ \Eprint {https://arxiv.org/abs/2003.07974} {arXiv:2003.07974} \BibitemShut {NoStop}%
\bibitem [{\citenamefont {Di~Biagio}\ \emph {et~al.}(2023)\citenamefont {Di~Biagio}, \citenamefont {Howl}, \citenamefont {Brukner}, \citenamefont {Rovelli},\ and\ \citenamefont {Christodoulou}}]{dibiagioRelativisticLocalityCan2023}%
  \BibitemOpen
  \bibfield  {author} {\bibinfo {author} {\bibfnamefont {A.}~\bibnamefont {Di~Biagio}}, \bibinfo {author} {\bibfnamefont {R.}~\bibnamefont {Howl}}, \bibinfo {author} {\bibfnamefont {{\v C}.}~\bibnamefont {Brukner}}, \bibinfo {author} {\bibfnamefont {C.}~\bibnamefont {Rovelli}},\ and\ \bibinfo {author} {\bibfnamefont {M.}~\bibnamefont {Christodoulou}},\ }\href {https://doi.org/10.48550/arXiv.2305.05645} {\bibinfo {title} {Relativistic locality can imply subsystem locality}} (\bibinfo {year} {2023}),\ \Eprint {https://arxiv.org/abs/2305.05645} {arXiv:2305.05645} \BibitemShut {NoStop}%
\bibitem [{\citenamefont {Polino}\ \emph {et~al.}(2022)\citenamefont {Polino}, \citenamefont {Polacchi}, \citenamefont {Poderini}, \citenamefont {Agresti}, \citenamefont {Carvacho}, \citenamefont {Sciarrino}, \citenamefont {Di~Biagio}, \citenamefont {Rovelli},\ and\ \citenamefont {Christodoulou}}]{polinoPhotonicImplementationQuantum2022}%
  \BibitemOpen
  \bibfield  {author} {\bibinfo {author} {\bibfnamefont {E.}~\bibnamefont {Polino}}, \bibinfo {author} {\bibfnamefont {B.}~\bibnamefont {Polacchi}}, \bibinfo {author} {\bibfnamefont {D.}~\bibnamefont {Poderini}}, \bibinfo {author} {\bibfnamefont {I.}~\bibnamefont {Agresti}}, \bibinfo {author} {\bibfnamefont {G.}~\bibnamefont {Carvacho}}, \bibinfo {author} {\bibfnamefont {F.}~\bibnamefont {Sciarrino}}, \bibinfo {author} {\bibfnamefont {A.}~\bibnamefont {Di~Biagio}}, \bibinfo {author} {\bibfnamefont {C.}~\bibnamefont {Rovelli}},\ and\ \bibinfo {author} {\bibfnamefont {M.}~\bibnamefont {Christodoulou}},\ }\href {https://doi.org/10.48550/arXiv.2207.01680} {\bibinfo {title} {Photonic {{Implementation}} of {{Quantum Gravity Simulator}}}} (\bibinfo {year} {2022}),\ \Eprint {https://arxiv.org/abs/2207.01680} {arXiv:2207.01680} \BibitemShut {NoStop}%
\end{thebibliography}%

\clearpage
\onecolumngrid
\section{Supplementary Material}
\emph{Mathematica notebooks---} 
We give an overview of the computation performed in the Mathematica notebooks published with this work. When descriptions differ, we will refer to the one used to achieve the relativistic corrections as \texttt{rel} and to the one used to calculate the entanglement of paths with uncertainty as \texttt{path}.The notebooks are available under \url{https://github.com/ofek-b/gme_superposition_geometries}. We first define three dimesionless quantities: $\phi = \frac{G m ^2}{\hbar \omega d}$, which is a measure for the effect of gravity in the system; $\epsilon = \alpha/d$, which is a measure of the quantum uncertainty of the particles' paths; and $\Omega = \omega d/c$, which is a measure for the effect of retardation in the system. When we write that we \emph{expand} quantities, it is meant that we perform a series expansion to second order in $\epsilon$ in both computations and to zeroth (second) order in $\Omega$ in \texttt{paths} (\texttt{rel}). Note that in \texttt{paths} we define $\xi=\beta/\alpha$ and so take second order in $\beta/d = \epsilon \xi$ automatically as well.

In \texttt{rel}, we find the retarded time $t_{ab}$ by solving a quadratic equation resulting from expanding its implicit definition $d_{ab}^2 =0$ with $d_{ab}$ defined in Eq.~\eqref{eq:dab}.

We then proceed to expand the action $S[q_1,q_2] = \int_0^t dt'(L_A + L_B + L_\text{G})$, a functional of general trajectories $q_1(t), q_2(t)$. We use the relativistic kinetic energy for the uncoupled part $L_A+L_B$ and the on-shell action Eq.~\eqref{eq:action_onshell} for $ L_\text{G}$. In \texttt{rel}, we introduce the bookkeeping variables ret and kin as follows to keep track of which relativistic contributions result from retardation (i.e., the finite-speed propagation of the change in spacetime's geometry) and which from kinetic energy and momenta gravitating (i.e., the correction to the sourcing of those changes predicted by GR). Note that this distinction is frame dependent and is not addressed in the main text, but will be used shortly.
% \begin{widetext}
\begin{align}
    (L_A+L_B)/\hbar \approx 
    \frac{q_1'(t)^2+q_2'(t)^2}{2 \omega }+\frac{\Omega ^2 \epsilon ^2 \left(q_1'(t)^4+q_2'(t)^4\right)}{8 \omega ^3}-\frac{2 \omega }{\Omega ^2 \epsilon ^2}, \nonumber
\end{align}
\begin{align}
L_\text{G}/\hbar \approx 
&\omega  \epsilon ^2 \phi  (q_1(t)+q_2(t))^2-\omega  \epsilon  \phi  (q_1(t)+q_2(t))+\omega  \phi  \nonumber\\ &
+\text{kin} \left(-\frac{4 \Omega ^2 \epsilon ^2 \phi  q_1'(t) q_2'(t)}{\omega }+\frac{3 \Omega ^2 \epsilon ^2 \phi  q_1'(t)^2}{2 \omega }+\frac{3 \Omega ^2 \epsilon ^2 \phi  q_2'(t)^2}{2 \omega }\right) \nonumber\\ &
+\text{ret}^2 \left(\frac{\Omega ^2 \epsilon  \phi  \left(q_1''(t)+q_2''(t)\right)}{2 \omega }-\frac{2 \Omega ^2 \epsilon ^2 \phi  q_1'(t) q_2'(t)}{\omega }+\frac{\Omega ^2 \epsilon ^2 \phi  q_1'(t)^2}{\omega }+\frac{\Omega ^2 \epsilon ^2 \phi  q_2'(t)^2}{\omega }\right)\nonumber\\ &
+\text{ret} \left(-\frac{\Omega ^2 \epsilon  \phi  \left(q_1''(t)+q_2''(t)\right)}{4 \omega }+\frac{\Omega ^2 \epsilon ^2 \phi  q_1'(t) q_2'(t)}{\omega }-\frac{\Omega ^2 \epsilon ^2 \phi  q_1'(t)^2}{2 \omega }-\frac{\Omega ^2 \epsilon ^2 \phi  q_2'(t)^2}{2 \omega }\right).
\nonumber
\end{align}
% \end{widetext}
The Lagrangian of \texttt{paths} (this of a Newtonian two-body problem) is achieved for kin, ret, $\Omega\rightarrow 0$. The Lagrangian of GR up to leading order in $1/c$ is achieved for taking both kin $=$ ret $=1$:
\begin{align}
    L_\text{G}/\hbar \approx &
    \omega  \epsilon ^2 \phi  (q_1(t)+q_2(t))^2-\omega  \epsilon  \phi  (q_1(t)+q_2(t))+\omega  \phi  
    +\frac{ \Omega ^2 \epsilon  \phi}{\omega} \left(\frac{q_1''(t) + q_2''(t)}{4} -5 \epsilon  q_1'(t) q_2'(t)+8 \epsilon  q_1'(t)^2 +2 \epsilon  q_2'(t)^2 \right) \nonumber
\end{align}

The classical trajectories beginning at $\mathbf{y}'_a$ and ending at $\mathbf{y}_a$ at time $t$ for the system are then calculated by solving the Euler-Lagrange equations resulting from the action. In \texttt{rel}, we first solve the non-relativistic equations of motion to achieve $\tilde{x}_a(t)$, and then plug in $q_a(t) = \tilde{x}_a(t) + \Omega c_a(t)$ to the relativistic equations of motion. We expand them and then solve for the correction $c_a(t)$. As it turns out, at the order of our expansion there is no relativistic correction to the classical trajectories. Note that as we have checked, the final result for the entanglement remains the same for both calculations if one takes the classical trajectories to be straight lines. This could be expected on order-of-magnitude grounds.

The trajectories are then plugged into the action and the wavefunction under the stationary phase approximation Eq.~\eqref{eq:psifi} is achieved after expanding the integrand and solving the integral. The wavefunction is a sum of polynomials multiplied by gaussians (in \texttt{rel} this sum has a single term, in \texttt{paths} two). Using a closed formula for the moments of the gaussian distribution we are able to perform the integral defining the generalized entanglement measure Eq.~\eqref{eq:def_ent} as a sum of expectation values of polynomials under a gaussian distribution (one in \texttt{rel}, sixteen in \texttt{paths}). Before solving the integral we expand the polynomials to an order double of the usual one, as ultimately a square root will be taken. We then arrive at our results Eqs.~\eqref{eq:ent_unified},~\eqref{eq:correction}. In the latter, we can set only one of kin and rel to one to achieve the different corrections from retardation and kinetic energy and momenta gravitating. We find that each time we get the correction reported in the main text multiplied by a numerical factor: $4/5$ and $1/5$ respectively.

\begin{figure}
    \centering
    \includegraphics{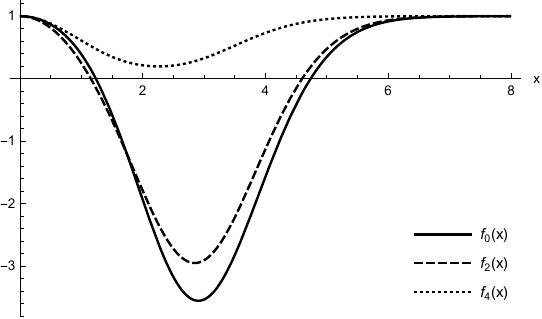}
    \caption{The functions from Eq.~\eqref{eq:ent_unified}. They are of order unit. The horizontal axis is $x=\beta/\alpha$.}
\end{figure}

\clearpage

   

\section*{Supplementary material}

# The Notebook paths.nb

In this notebook we arrive at the result in Eq. (11), which is the generalized entanglement measure for the intermediate case of path superposition with non-negligible gaussian width. The calculation proceeds as follows: (1) defining the non-relativistic action, (2) finding the classical solutions of the system, (3) computing the gaussian integrals in Eq. (9), (4) calculating for this wavefunction the generalized entanglement measure [Swain et al, Phys. Rev. A 105, 052441 (2022)] defined in Eq. (28), to get Eq. (11).

```
<< VariationalMethods`

$Assumptions = {m > 0, d > 0, ω > 0, ξ > 0, t > 0, ϕ > 0 , ϵ > 0 , Ω > 0 ,
    ti > 0, y1 ∈ Reals, y2 ∈ Reals, y1p ∈ Reals, y2p ∈ Reals, 4 ϵ^2 ϕ < 1,
    trp ≥ 0, ξ1 ∈ Reals, ξ2 ∈ Reals, s1i ∈ Reals, s2i ∈ Reals, s1j ∈ Reals,
    s2j ∈ Reals, s1k ∈ Reals, s2k ∈ Reals, s1l ∈ Reals, s2l ∈ Reals};

(*defining the dimensionless quantities ϵ,
ϕ and Ω which fully define the problem.*)
ℏ = ϵ^2 d^2 m ω;
(*ϵ = α/d is a measure of the "quantumness" of the particles*)
G = ϕ ℏ d ω / (m^2 ) ;
(*ϕ = Gm²/ℏωd is a measure of the effect of gravity on the system*)
c = d ω / (Ω ) ;
(*Ω = ωd/c is a measure of the effect of retardation in the system*)

α = Sqrt[ℏ / (m ω)]; (*the width of the gaussian*)

(*a function calculating an integral of the form fPre Exp[fExp],
where fPre is a polynomial and fExp a 2nd order polynomial. Taken
   from https://mathematica.stackexchange.com/a/6846.*)
gaussMoment[fPre_, fExp_, vars_] := Module[{coeff, dist, ai, μ, norm},
   coeff = CoefficientArrays[fExp, vars, "Symmetric" → True];
   ai = Inverse[2 coeff[[3]]];
   μ = -ai.coeff[[2]];
   dist = MultinormalDistribution[μ, -ai];
   norm = 1 / PDF[dist, vars] /. Thread[vars → μ];
   norm Exp[1 / 2 coeff[[2]].μ + coeff[[1]]] ×
    Distribute@Expectation[fPre, vars ≈ dist]]

(*functions replacing coefficients of given polynomials with
 single symbols. This makes mathematica work much faster.*)
myMatrix[symb_, d1_, d2_, d3_, d4_] :=
   Table[Subscript[symb, i1, i2, i3, i4], {i1, d1}, {i2, d2}, {i3, d3}, {i4, d4}];
symbolize[poly_, vars2_, symb_] :=
```

```
  Module[{lst, flst, symbolized, dims, symbCoeffs},
   lst = CoefficientList[poly, vars2];
   dims = Dimensions[lst];
   symbCoeffs = myMatrix[symb, dims[[1]], dims[[2]], dims[[3]], dims[[4]]];
   Do[
    If[lst[[i1, i2, i3, i4]] == 0, symbCoeffs[[i1, i2, i3, i4]] = 0];
    , {i1, 1, dims[[1]]}, {i2, 1, dims[[2]]}, {i3, 1, dims[[3]]}, {i4, 1, dims[[4]]}];
   symbolized = Internal`FromCoefficientList[symbCoeffs, vars2];
   symbolized
  ];

(*a function wrapping gaussMoment,
replacing coefficients with symbols before integrating.*)
myGauss[fPre_, fExp_, vars1_] :=
 Module[{a0exp, a0pre, aexp, apre, sympre, symexp, res},
  a0exp = CoefficientList[fExp, vars];
  symexp = symbolize[fExp, vars, aexp];
  a0pre = CoefficientList[fPre, vars];
  sympre = symbolize[fPre, vars, apre];
  res = gaussMoment[ sympre , symexp, vars1] /.
    {aexp → a0exp, apre → a0pre, Subscript → Part};
  res
 ]

(*operations on complex numbers done in a
 custom way which for some reason works faster.*)
myRe[var_] := Module[{var1, var1C, re, im},
  var1 = var // ComplexExpand // Evaluate;
  var1C = var1 // Conjugate // Refine;
  re = (var1 + var1C) / 2 // Simplify;
  re
 ]

myIm[var_] := Module[{var1, var1C, re, im},
  var1 = var // ComplexExpand // Evaluate;
  var1C = var1 // Conjugate // Refine;
  im = -i (var1 - var1C) / 2 // Simplify;
  im
 ]

myConj[var_] := Module[{var1, var1C, re, im},
  var1 = var // ComplexExpand // Evaluate;
  var1C = var1 // Conjugate // Refine;
  var1C
 ]

(*the series expansion to second order in ϵ and
```

```
  zeroth order in Ω. The second function is for quantities 
  which will be taken with a square root later.*) 
mySeries[expr_] := Series[ expr, {Ω, 0, 0}, {ϵ, 0, 2}] // Normal;
mySeriesSq[expr_] := Series[expr, {Ω, 0, 0}, {ϵ, 0, 4}] // Normal;
```

## Action

We define action of the gravitational interaction: SGhb = $S_G / \hbar$. This is the Newtonian potential, as we work in the non-relativistic limit in this calculation. The functions q1, q2 are the variables for the lagrangian and action.

*In[ ]:=* 
```
LG12 = (G m^2 / 2) / (d + α q1[ti] + α q2[ti]) / ℏ    // mySeries // Simplify ;
LGhb = LG12 + (LG12 /. {q1 → q2, q2 → q1}) // mySeries // Simplify
SGhb = Integrate[ LGhb, {ti, 0, t}] ;
```

*Out[ ]=*

$$\phi \, \omega \left(1 - \epsilon \, (q1[ti] + q2[ti]) + \epsilon^2 \, (q1[ti] + q2[ti])^2\right)$$

The free, uncoupled action: Sunc = $\Sigma_{a=1,2} \, S_a / \hbar$:

*In[ ]:=*
```
Lunc = ( m (α q1'[ti])^2 / 2 + m (α q2'[ti])^2 / 2) / ℏ // mySeries // Simplify
Sunc = Integrate[ Lunc, {ti, 0, t}] ;
```

*Out[ ]=*

$$\frac{q1'[ti]^2 + q2'[ti]^2}{2 \, \omega}$$

Now that we have defined the action, we proceed to finding the solution to the classical equations of motion, which we will use for the stationary-phase approximation in Eq. (9).

## Classical Solutions

We find the classical trajectories of the particles for initial locations y1p, y2p = $y'_1 / \alpha$, $y'_2 / \alpha$ and final locations y1, y2 = $y_1 / \alpha$, $y_2 / \alpha$. Here, the y's are in units of $\alpha$, the gaussian's width. We do it by solving the system under the second order approximation in $\epsilon$. The problem becomes two harmonic oscillators with some frequency trp*$\omega$, which are coupled by another spring, which is gravity. The variable trp is just a bookkeeping variable which will be taken to zero to achieve the trajectories of the two otherwise free particles coupled by gravity.

*In[ ]:=*
```
L[ti] := (LGhb + Lunc) ;
L[ti] // Simplify
```

*Out[ ]=*

$$\phi \, \omega \left(1 - \epsilon \, (q1[ti] + q2[ti]) + \epsilon^2 \, (q1[ti] + q2[ti])^2\right) + \frac{q1'[ti]^2 + q2'[ti]^2}{2 \, \omega}$$

*In[ ]:=* `eqs = EulerEquations[L[ti], {q1[ti], q2[ti]}, ti] //`
      `Expand (*the classical equations of motion*)`

*Out[ ]=*
$$\left\{-\epsilon \phi \omega + 2 \epsilon^2 \phi \omega \, q1[ti] + 2 \epsilon^2 \phi \omega \, q2[ti] - \frac{q1''[ti]}{\omega} == 0,\right.$$
$$\left.-\epsilon \phi \omega + 2 \epsilon^2 \phi \omega \, q1[ti] + 2 \epsilon^2 \phi \omega \, q2[ti] - \frac{q2''[ti]}{\omega} == 0\right\}$$

*In[ ]:=* `sol = DSolve[{eqs〚1〛, eqs〚2〛, q1[0] == y1p, q2[0] == y2p, q1[t] == y1, q2[t] == y2},`
      `{q1[ti], q2[ti]}, {ti}] // Simplify;(*their solution*)`

*In[ ]:=*
   `x1[time_] := q1[ti] /. sol〚1〛 /. ti → time;`
   `x2[time_] := q2[ti] /. sol〚1〛 /. ti → time;`
   `x1[ti]`
   `x2[ti]`

*Out[ ]=*
$$\frac{1}{4\left(-1+e^{4t\epsilon\sqrt{\phi}\,\omega}\right)t\epsilon}$$
$$e^{-2ti\epsilon\sqrt{\phi}\,\omega}\left(-e^{2t\epsilon\sqrt{\phi}\,\omega}\,t\,(-1+2y1\epsilon+2y2\epsilon) + e^{2(t+2ti)\epsilon\sqrt{\phi}\,\omega}\,t\,(-1+2y1\epsilon+2y2\epsilon) + \right.$$
$$e^{4t\epsilon\sqrt{\phi}\,\omega}\,t\,(-1+2y1p\,\epsilon+2y2p\,\epsilon) - e^{4ti\epsilon\sqrt{\phi}\,\omega}\,t\,(-1+2y1p\,\epsilon+2y2p\,\epsilon) -$$
$$e^{2ti\epsilon\sqrt{\phi}\,\omega}\,(2ti\,(y1-y1p-y2+y2p)\,\epsilon + t\,(1+2y1p\,\epsilon-2y2p\,\epsilon)) +$$
$$\left.e^{2(2t+ti)\epsilon\sqrt{\phi}\,\omega}\,(2ti\,(y1-y1p-y2+y2p)\,\epsilon + t\,(1+2y1p\,\epsilon-2y2p\,\epsilon))\right)$$

*Out[ ]=*
$$\frac{1}{4\left(-1+e^{4t\epsilon\sqrt{\phi}\,\omega}\right)t\epsilon}$$
$$e^{-2ti\epsilon\sqrt{\phi}\,\omega}\left(-e^{2t\epsilon\sqrt{\phi}\,\omega}\,t\,(-1+2y1\epsilon+2y2\epsilon) + e^{2(t+2ti)\epsilon\sqrt{\phi}\,\omega}\,t\,(-1+2y1\epsilon+2y2\epsilon) + \right.$$
$$e^{4t\epsilon\sqrt{\phi}\,\omega}\,t\,(-1+2y1p\,\epsilon+2y2p\,\epsilon) - e^{4ti\epsilon\sqrt{\phi}\,\omega}\,t\,(-1+2y1p\,\epsilon+2y2p\,\epsilon) +$$
$$e^{2ti\epsilon\sqrt{\phi}\,\omega}\,(2ti\,(y1-y1p-y2+y2p)\,\epsilon + t\,(-1+2y1p\,\epsilon-2y2p\,\epsilon)) +$$
$$\left.e^{2(2t+ti)\epsilon\sqrt{\phi}\,\omega}\,(2ti\,(-y1+y1p+y2-y2p)\,\epsilon + t\,(1-2y1p\,\epsilon+2y2p\,\epsilon))\right)$$

Sanity check: initial and final locations

*In[ ]:=* `x1[0] // mySeries // Simplify`
   `x2[0] // mySeries // Simplify`
   `x1[t] // mySeries // Simplify`
   `x2[t] // mySeries // Simplify`

*Out[ ]=*
   y1p

*Out[ ]=*
   y2p

*Out[ ]=*
   y1

*Out[ ]=*
   y2

sanity check: check that the solution actually solves the classical equation

```
In[ ]:= L[ti] := LGhb + Lunc;
        EulerEquations[L[ti], {q1[ti], q2[ti]}, ti] /. {q1 → x1, q2 → x2} //
          mySeries // Simplify
```

*Out[ ]=*

{True, True}

---

## Wavefunction

We would now like to compute the wavefunction $\tilde{\psi}^{fi}(y1, y2)$ as given in Eq. (9). We start with achieving the expression for the on-shell ( = stationary-phase approximated) action $\tilde{S}_a + \tilde{S}_G$, here Shb.

```
In[ ]:= Lhb = LGhb + Lunc /. {q1 → x1, q2 → x2} // mySeries // Simplify;
        Shb1 = Integrate[Lhb , {ti, 0, t}] ;
        Shb =
         Shb1 - (Shb1 /. {y1 → 0, y2 → 0, y1p → 0, y2p → 0}) // Expand // Refine // Simplify
```

*Out[ ]=*

$$\frac{1}{6\,t\,\omega} \left(3\,y2^2 - 6\,y2\,y2p + 3\,y2p^2 - 3\,t^2\,y2\,\epsilon\,\phi\,\omega^2 - \right.$$
$$3\,t^2\,y2p\,\epsilon\,\phi\,\omega^2 + 2\,t^2\,y2^2\,\epsilon^2\,\phi\,\omega^2 + 2\,t^2\,y2\,y2p\,\epsilon^2\,\phi\,\omega^2 + 2\,t^2\,y2p^2\,\epsilon^2\,\phi\,\omega^2 +$$
$$t^2\,y1\,\epsilon\,(-3 + 4\,y2\,\epsilon + 2\,y2p\,\epsilon)\,\phi\,\omega^2 + t^2\,y1p\,\epsilon\,(-3 + 2\,y2\,\epsilon + 4\,y2p\,\epsilon)\,\phi\,\omega^2 +$$
$$\left. 2\,y1\,y1p\,\left(-3 + t^2\,\epsilon^2\,\phi\,\omega^2\right) + y1^2\,\left(3 + 2\,t^2\,\epsilon^2\,\phi\,\omega^2\right) + y1p^2\,\left(3 + 2\,t^2\,\epsilon^2\,\phi\,\omega^2\right)\right)$$

We now solve the y1p and y2p integrals while also performing the approximations. We denote for any function f the parts fExp and fPre such that $f(x) = fPre(x) \times e^{fExp(x)}$.

The initial state is a sum of four gaussians with $\xi 1, \xi 2 = \pm \beta/\alpha$. We therefore perform one calculation (the variable summand) and then use it four times with the appropriate replacements of $\xi 1, \xi 2$.

```
In[ ]:= vars = {y1, y2, y1p, y2p};
        summand = Exp[ - (y1p - ξ1)^2 / 2 - (y2p - ξ2)^2 / 2] *
          (ⅈ / ((-ⅈ + ω t) (3 + 3 ⅈ ω t + 4 ϵ² ϕ (ω t)²)))^(-1/2);
        (*four of these make up the initial state. prefactor
         added for a simpler expression later*)

        wf4vars = summand Exp[ⅈ Shb] // mySeriesSq // Simplify;
        (*the integrand in Eq. (9).*)
        wf4varsExp = Exponent[wf4vars, ⅇ] // mySeries // Simplify;
        wf4varsPre = wf4vars Exp[-wf4varsExp] // mySeriesSq // Simplify;
        wf = myGauss[ wf4varsPre , wf4varsExp, {y1p, y2p}] // mySeriesSq // Simplify;
        (*This is $\tilde{\psi}^{fi}$(y1, y2)*)
        wfExp = Exponent[wf, ⅇ] // mySeries // Simplify;
        wfPre = wf Exp[-wfExp] // mySeriesSq // Simplify;
        wfExpC = wfExp // myConj // Simplify; (*complex conjugate*)
        wfPreC = wfPre // myConj // Simplify;
```

calculate the norm squared normsq of the state, integrating each separate exponent separately using our myGauss function:

```mathematica
wfExp0 = {wfExp /. {ξ1 → ξ, ξ2 → ξ}, wfExp /. {ξ1 → -ξ, ξ2 → ξ},
    wfExp /. {ξ1 → ξ, ξ2 → -ξ}, wfExp /. {ξ1 → -ξ, ξ2 → -ξ}};
wfExpC0 = {wfExpC /. {ξ1 → ξ, ξ2 → ξ}, wfExpC /. {ξ1 → -ξ, ξ2 → ξ},
    wfExpC /. {ξ1 → ξ, ξ2 → -ξ}, wfExpC /. {ξ1 → -ξ, ξ2 → -ξ}};
wfPre0 = {wfPre /. {ξ1 → ξ, ξ2 → ξ}, wfPre /. {ξ1 → -ξ, ξ2 → ξ},
    wfPre /. {ξ1 → ξ, ξ2 → -ξ}, wfPre /. {ξ1 → -ξ, ξ2 → -ξ}};
wfPreC0 = {wfPreC /. {ξ1 → ξ, ξ2 → ξ}, wfPreC /. {ξ1 → -ξ, ξ2 → ξ},
    wfPreC /. {ξ1 → ξ, ξ2 → -ξ}, wfPreC /. {ξ1 → -ξ, ξ2 → -ξ}};

normsq = 0;
Do[
 preee = wfPreC0[[i]] × wfPre0[[j]] // mySeriesSq;
 preee = preee // Simplify;
 res = myGauss[preee, wfExpC0[[i]] + wfExp0[[j]], {y1, y2}];
 res = res // mySeriesSq;
 res = res // Simplify;
 If[i ≠ j, res = 2 myRe[res];];
 normsq += res;
 , {i, 1, Length[wfExp0]}, {j, i, Length[wfExp0]}]
normsq = normsq // Simplify;
```

## Generalized Entanglement Measure

we now calculate the entanglement measure in Eq. (28) after replacing the norm square in the formula by multiplication with the conjugate:

$$\left| \int \psi(y1, y2) \psi^*(y'1, y2) \, dy2 \right|^2 = \int \psi^*(y1, y2) \psi(y'1, y2) \psi(y1, y'2) \psi^*(y'1, y'2) \, dy2 \, dy'2$$

each $\psi$ contains 4 elements, so we have 16 elements in the integrand. We calculate only once (variable summand2), and then sum over the 16 possible choices of the signs in the $\xi$'s.

```mathematica
(*each of the four multipicants inside the integral above is
  slightly different in where is the ' and whether it is conjugate*)
wfPre1p = wfPre /. y1 → y1p;
wfExp1p = wfExp /. y1 → y1p;
wfPre2p = wfPre /. y2 → y2p;
wfExp2p = wfExp /. y2 → y2p;
wfPreC12p = wfPreC /. {y1 → y1p, y2 → y2p};
wfExpC12p = wfExpC /. {y1 → y1p, y2 → y2p};

(*we again get to the form f(x)=
  fPre(x)*e^fExp(x) before we calculate the integral using myGauss*)
thisPre = (wfPre1p /. {ξ1 → s1i ξ, ξ2 → s2i ξ})
    (wfPreC /. {ξ1 → s1j ξ, ξ2 → s2j ξ}) (wfPreC12p /. {ξ1 → s1k ξ, ξ2 → s2k ξ})
    (wfPre2p /. {ξ1 → s1l ξ, ξ2 → s2l ξ}) // mySeriesSq // Simplify;
Print["pre series"];
thisExp = (wfExp1p /. {ξ1 → s1i ξ, ξ2 → s2i ξ}) +
    (wfExpC /. {ξ1 → s1j ξ, ξ2 → s2j ξ}) + (wfExpC12p /. {ξ1 → s1k ξ, ξ2 → s2k ξ}) +
    (wfExp2p /. {ξ1 → s1l ξ, ξ2 → s2l ξ}) // Simplify;
Print["exp simp"];
summand2 = myGauss[thisPre, thisExp, {y1, y2, y1p, y2p}];
Print["int"];
summand2 = summand2 / normsq^2 // mySeriesSq // Simplify; (*normalize*)
Print["int ser"];
summand2 = summand2 // Simplify;
Print["int simp"];
```

pre series

exp simp

int

int ser

int simp

now sum over the 16 elements:

*In[ ]:=* ```mathematica
sum = Sum[ summand2, {s1i, {-1, 1}}, {s2i, {-1, 1}}, {s1j, {-1, 1}}, {s2j, {-1, 1}},
    {s1k, {-1, 1}}, {s2k, {-1, 1}}, {s1l, {-1, 1}}, {s2l, {-1, 1}}] // Simplify;
```

finally, the generalized entanglement measure (Eq. 28) is:

*In[ ]:=* ```mathematica
gem = Sqrt[2 (1 - sum)]  // Expand // Simplify
```

*Out[ ]=*

$\left(1 \Big/ \left(3 \left(1 + e^{\xi^2}\right)\right)\right) 2 t \epsilon^2 \phi \omega$
$\sqrt{\left(9 + 3 t^2 \left(1 - 3 \xi^2\right) \omega^2 + t^4 \left(1 - 2 \xi^2\right)^2 \omega^4 + e^{2\xi^2} \left(9 + 36 \xi^4 + 3 t^2 \omega^2 + t^4 \omega^4 + 9 \xi^2 \left(4 + t^2 \omega^2\right)\right) + 2 e^{\xi^2} \left(9 + 3 t^2 \omega^2 - 9 t^2 \xi^4 \omega^2 + t^4 \omega^4 - 2 \xi^2 \left(-9 + t^4 \omega^4\right)\right)\right)}$

find a nicer representation (Eq. 11) with order 1 functions (Eq. 29):

*In[ ]:=* ```mathematica
coefflist = (gem / (2 t ϵ^2 ϕ ω))^2 /. {ξ → x, ω → 1} // CoefficientList[#, t] & ;
```

*In[ ]:=* `f0 = coefflist[[1]] - 4 x^2 (1 + x^2) // Simplify`

*Out[ ]=*

$$\left(1 + e^{2x^2} - 4x^2 - 4x^4 + e^{x^2}\left(2 - 4x^2 - 8x^4\right)\right) \Big/ \left(1 + e^{x^2}\right)^2$$

*In[ ]:=* `f2 = 3 coefflist[[3]] - 3 x^2 // Simplify`

*Out[ ]=*

$$\left(1 + e^{2x^2} - 6x^2 + e^{x^2}\left(2 - 6x^2 - 6x^4\right)\right) \Big/ \left(1 + e^{x^2}\right)^2$$

*In[ ]:=* `f4 = 9 coefflist[[5]] // Simplify`

*Out[ ]=*

$$\left(1 + e^{x^2} - 2x^2\right)^2 \Big/ \left(1 + e^{x^2}\right)^2$$

```
gem^2 -
   ( 2 t e^2 ϕ ω Sqrt[f0 + 4 x^2 (1 + x^2) + f2 (ω t)^2 / 3 + x^2 (ω t)^2 + f4 (ω t)^4 / 9])^
    2 /. ξ → x // FullSimplify
 (*a sanity check: this should be zero if Eq. (11) really
    equals to our result from gem above.*)
```

*Out[ ]=*

0

Plot the functions:

```
Plot[{((1 + e^(2 x^2) - 4 x^2 - 4 x^4 + e^(x^2) (2 - 4 x^2 - 8 x^4)) / (1 + e^(x^2))^2 /. x → xx / 2,
    (1 + e^(2 x^2) - 6 x^2 + e^(x^2) (2 - 6 x^2 - 6 x^4)) / (1 + e^(x^2))^2 /. x → xx / 2,
    (1 + e^(x^2) - 2 x^2)^2 / (1 + e^(x^2))^2 /. x → xx / 2}, {xx, 0, 8},
  PlotLegends → Placed[{"f₀(x)", "f₂(x)", "f₄(x)"}, {Right, Bottom}],
  PlotTheme → "Monochrome", AxesLabel → {"x"}]
 (*the x is divided by two because in the paper the distance
    between the paths is β and in this notebook it is 2β.*)
```

*Out[ ]=*

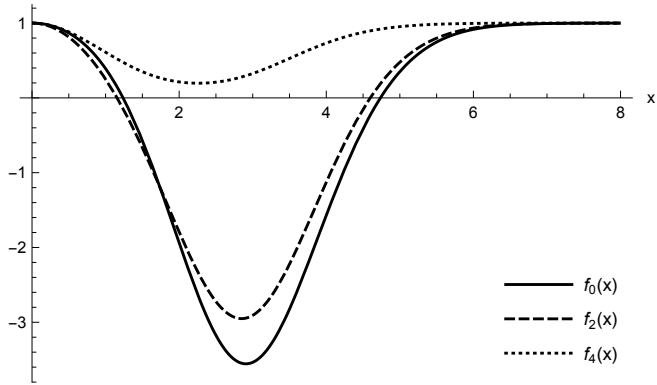

# The Notebook rel.nb

In this notebook we arrive at the result in Eq. (30) (based on which Eq, (16) was also found), which is the generalized entanglement measure to leading order in 1/c for the case of two gaussians. The calculation proceeds as follows: (1) solving the implicit definition of the retarded time from the paragraph after Eq. (24) to the desired order, (2) defining the relativistic action Eq. (21) , (2) finding the classical solutions of the system, (3) computing the gaussian integrals of which Eq. (9) consists, (4) calculating for this wavefunction the generalized entanglement measure [Swain et al, Phys. Rev. A 105, 052441 (2022)] defined in Eq. (28), to get Eq. (30).

```
In[ ]:=  << VariationalMethods`

$Assumptions = {m > 0, d > 0, ω > 0, t > 0, ti > 0,
   tii > 0, ϕ > 0 , ϵ > 0 , Ω > 0 , y1 ∈ Reals, y2 ∈ Reals, y1p ∈ Reals,
   y2p ∈ Reals, dtN > 0, 4 ϵ^2 ϕ < 1, trp ≥ 0, kin ≥ 0, ret ≥ 0};

(*defining the dimensionless quantities ϵ,
ϕ and Ω which fully define the problem.*)
ℏ = ϵ^2 d^2 m ω;
(*ϵ = α/d is a measure of the "quantumness" of the particles*)
G = ϕ ℏ d ω / (m^2 ) ;
(*ϕ = Gm²/ℏωd is a measure of the effect of gravity on the system*)
c = d ω / (Ω );
(*Ω = ωd/c is a measure of the effect of retardation in the system*)

α = Sqrt[ℏ / (m ω)]; (*the width of the gaussian*)

(*a function calculating an integral of the form fPre Exp[fExp],
where fPre is a polynomial and fExp a 2nd order polynomial. Taken
  from https://mathematica.stackexchange.com/a/6846.*)
gaussMoment[fPre_, fExp_, vars_] := Module[{coeff, dist, ai, μ, norm},
   coeff = CoefficientArrays[fExp, vars, "Symmetric" → True];
   ai = Inverse[2 coeff〚3〛];
   μ = - ai.coeff〚2〛;
   dist = MultinormalDistribution[μ, -ai];
   norm = 1 / PDF[dist, vars] /. Thread[vars → μ];
   norm Exp[1 / 2 coeff〚2〛.μ + coeff〚1〛] ×
    Distribute@Expectation[fPre, vars ≈ dist]]

(*functions replacing coefficients of given polynomials with
 single symbols. This makes mathematica work much faster.*)
myMatrix[symb_, d1_, d2_, d3_, d4_] :=
   Table[Subscript[symb, i1, i2, i3, i4], {i1, d1}, {i2, d2}, {i3, d3}, {i4, d4}];
symbolize[poly_, vars2_, symb_] :=
```

```mathematica
  Module[{lst, flst, symbolized, dims, symbCoeffs},
   lst = CoefficientList[poly, vars2];
   dims = Dimensions[lst];
   symbCoeffs = myMatrix[symb, dims[[1]], dims[[2]], dims[[3]], dims[[4]]];
   Do[
    If[lst[[i1, i2, i3, i4]] ⩵ 0, symbCoeffs[[i1, i2, i3, i4]] = 0];
    , {i1, 1, dims[[1]]}, {i2, 1, dims[[2]]}, {i3, 1, dims[[3]]}, {i4, 1, dims[[4]]}];
   
   symbolized = Internal`FromCoefficientList[symbCoeffs, vars2];
   symbolized
   ];

(*a function wrapping gaussMoment,
replacing coefficients with symbols before integrating.*)
myGauss[fPre_, fExp_, vars1_] :=
 Module[{a0exp, a0pre, aexp, apre, sympre, symexp, res},
  a0exp = CoefficientList[fExp, vars];
  symexp = symbolize[fExp, vars, aexp];
  a0pre = CoefficientList[fPre, vars];
  sympre = symbolize[fPre, vars, apre];
  res = gaussMoment[ sympre , symexp, vars1] /.
     {aexp → a0exp, apre → a0pre, Subscript → Part};
  res
 ]

(*operations on complex numbers done in a custom way
 which for some reason works faster, at least in this code.*)
myRe[var_] := Module[{var1, var1C, re, im},
  var1 = var // ComplexExpand // Evaluate;
  var1C = var1 // Conjugate // Refine;
  re = (var1 + var1C) / 2 // Simplify;
  re
 ]

myIm[var_] := Module[{var1, var1C, re, im},
  var1 = var // ComplexExpand // Evaluate;
  var1C = var1 // Conjugate // Refine;
  im = -ⅈ (var1 - var1C) / 2 // Simplify;
  im
 ]

myConj[var_] := Module[{var1, var1C, re, im},
  var1 = var // ComplexExpand // Evaluate;
  var1C = var1 // Conjugate // Refine;
  var1C
 ]
```

```
(*the series expansion to second order in ϵ and
 second order in Ω. The second function is for quantities
 which will be taken with a square root later.*)
mySeries[expr_] := Series[ expr, {Ω, 0, 2}, {ϵ, 0, 2}] // Normal;
mySeriesSq[expr_] := Series[expr, {Ω, 0, 4}, {ϵ, 0, 4}] // Normal;
```

## Find the retarded time to relevant order

Expand the definition of the retarded time $t_{ab}$ to second order in $\Omega = \omega d/c$ and solve:

```
dtNequals = 1 + (α / d ) q1[ti] + (α / d ) q2[ti - dtN d / c] // Simplify;
(*the implicit expression of the retarded time difference from the paragraph
 after Eq.(24). We work with normalized time, in units of d/c.*)
dtNequalsSer = dtNequals // mySeries // Expand (*expand the expression*)
lst = Roots[dtNequalsSer - dtN ⩵ 0 , dtN] ;
(*solve the polynomial equation obtained by equating the
 retarded time difference to the expanded implicit expression.*)
If[Length[lst〚1〛] < 2, lst = List[lst]];

(*loop to find the right solution dtNsol,
i.e. positive and 1 in the non-relativistic limit*)
For[i = 1, i ≤ Length[lst], i++,
  Print["checking solution number ", i];
  this = lst〚i〛 /. Equal → List;
  sol = this〚2〛 // Simplify;
  Print[sol];
  ldng = Series[sol, ϵ → 0, Ω → 0] // Normal // Simplify;
  If[Not[ldng ⩵ 1], Continue[]];
  If[Not[sol > 0], Continue[]];
  Print["Found"];
  dtNsol = sol;
  Break[];
  ];
```

*Out[•]=*

$$1 + \epsilon\, q1[ti] + \epsilon\, q2[ti] - \frac{dtN\, \epsilon\, \Omega\, q2'[ti]}{\omega} + \frac{dtN^2\, \epsilon\, \Omega^2\, q2''[ti]}{2\, \omega^2}$$

checking solution number 1

$$\frac{\omega\left(\omega + \epsilon\, \Omega\, q2'[ti] - \sqrt{(\omega + \epsilon\, \Omega\, q2'[ti])^2 - 2\, \epsilon\, \Omega^2\, (1 + \epsilon\, q1[ti] + \epsilon\, q2[ti])\, q2''[ti]}\right)}{\epsilon\, \Omega^2\, q2''[ti]}$$

Found

## Action

The action of the gravitational interaction from Eq. (21): SGhb = $S_G / \hbar$ . Below the retarded quanti-

ties in it are defined. ret and kin are bookkeeping variables to keep track of contributions from the retardation part and kinetic part of the GR correction.

```
In[ ]:= tab = ti - ret dtNsol d / c; (*the retarded time*)
        dab = d dtNsol; (*the retarded distance*)
        γ[v_] := (1 - v^2 / c^2) ^ (-1 / 2) (*Lorentz factor*)
        VV[ti_, xa_, xaDot_, xb_, xbDot_, a_, b_] :=
          c^4 γ[α xaDot[tab]] × γ[α xbDot[ti]] ( (1 - α xaDot[tab] α xbDot[ti] / c^2) ^2 -
            1
            - (1 - α^2 xaDot[tab]^2 / c^2) (1 - α^2 xbDot[ti]^2 / c^2) )
            2
        LG12 =
          (G m^2 ((Series[VV[ti, q1, q1', q2, q2', a, b] / c^4, Ω → 0] // Normal) (1 - kin) +
                kin VV[ti, q1, q1', q2, q2', a, b] / c^4)) /
              ((Series[dab - dab * (-α q1'[tab]) / c, Ω → 0] // Normal) (1 - ret) +
                ret (dab - dab * (-α q1'[tab]) / c))   / ℏ // Simplify ;
        LGhb = (LG12 + (LG12 /. {q1 → q2, q2 → q1}) ) // mySeries //
          Simplify (*the lagrangian of the on-shell gravitational action,
          we sum over the potential induced by 1 on 2 and 2 on 1.*)
```

Out[ ]=

$$\phi \omega - \epsilon \phi \omega \, (q1[ti] + q2[ti]) + \epsilon^2 \phi \omega \, (q1[ti] + q2[ti])^2 + \frac{1}{4\omega} \epsilon \phi \Omega^2$$
$$\left(2 \, (3\,\text{kin} + \text{ret}\,(-1+2\,\text{ret}))\, \epsilon\, q1'[ti]^2 - 4\, (4\,\text{kin} + \text{ret}\,(-1+2\,\text{ret}))\, \epsilon\, q1'[ti]\, q2'[ti] + 2\, (3\,\text{kin} + \text{ret}\,(-1+2\,\text{ret}))\, \epsilon\, q2'[ti]^2 + \text{ret}\,(-1+2\,\text{ret})\,(q1''[ti] + q2''[ti])\right)$$

the full gravity lagrangian for kin=ret=1:

```
In[ ]:= LGhb /. {kin → 1, ret → 1} // Simplify
```

Out[ ]=

$$\phi \omega - \epsilon \phi \omega \, (q1[ti] + q2[ti]) + \epsilon^2 \phi \omega \, (q1[ti] + q2[ti])^2 + $$
$$\frac{\epsilon \phi \Omega^2 \, \left(8 \, \epsilon \, q1'[ti]^2 - 20\, \epsilon\, q1'[ti]\, q2'[ti] + 8\, \epsilon\, q2'[ti]^2 + q1''[ti] + q2''[ti]\right)}{4\omega}$$

The free, uncoupled action: Sunc = $\Sigma_{a=1,2} S_a / \hbar$, its lagrangian:

```
In[ ]:= Lunc = ( - m c^2 / γ[α q1'[ti]] - m c^2 / γ[α q2'[ti]] ) / ℏ // mySeries // Simplify
```

Out[ ]=

$$- \frac{2\omega}{\epsilon^2 \Omega^2} + \frac{q1'[ti]^2 + q2'[ti]^2}{2\omega} + \frac{\epsilon^2 \Omega^2 \, \left(q1'[ti]^4 + q2'[ti]^4\right)}{8\omega^3}$$

# Classical Solutions

We find the classical trajectories of the particles for initial locations y1p, y2p = $y'_1 / \alpha$, $y'_2 / \alpha$ and final locations y1, y2 = $y_1 / \alpha$, $y_2 / \alpha$. Here, the y's are in units of $\alpha$. We do it by solving the system under the second order approximation in $\epsilon$ and in $\Omega$. The problem becomes two harmonic oscillators with some frequency trp×$\omega$, which are coupled by another spring, which is gravity. The variable trp is just a bookkeeping variable which will be taken to zero to achieve the trajectories of the two otherwise free particles coupled by gravity.

*In[ ]:=*
```
L[ti] := (LGhb + Lunc) ;
L[ti] // Simplify
```

*Out[ ]=*

$$\phi\,\omega - \frac{2\,\omega}{\epsilon^2\,\Omega^2} - \epsilon\,\phi\,\omega\,(q1[ti] + q2[ti]) + \epsilon^2\,\phi\,\omega\,(q1[ti] + q2[ti])^2 +$$
$$\frac{q1'[ti]^2 + q2'[ti]^2}{2\,\omega} + \frac{\epsilon^2\,\Omega^2\,\left(q1'[ti]^4 + q2'[ti]^4\right)}{8\,\omega^3} + \frac{1}{4\,\omega}\,\epsilon\,\phi\,\Omega^2$$
$$\left(2\,(3\,kin + ret\,(-1 + 2\,ret))\,\epsilon\,q1'[ti]^2 - 4\,(4\,kin + ret\,(-1 + 2\,ret))\,\epsilon\,q1'[ti]\,q2'[ti] + \right.$$
$$\left. 2\,(3\,kin + ret\,(-1 + 2\,ret))\,\epsilon\,q2'[ti]^2 + ret\,(-1 + 2\,ret)\,(q1''[ti] + q2''[ti])\right)$$

*In[ ]:=*
```
eqs = EulerEquations[L[ti], {q1[ti], q2[ti]}, ti] // Expand;
(*the classical equations of motion*)
Collect[eqs, Ω]
```

*Out[ ]=*

$$\left\{-\epsilon\,\phi\,\omega + 2\,\epsilon^2\,\phi\,\omega\,q1[ti] + 2\,\epsilon^2\,\phi\,\omega\,q2[ti] - \right.$$
$$\frac{q1''[ti]}{\omega} + \Omega^2\left(-\frac{3\,kin\,\epsilon^2\,\phi\,q1''[ti]}{\omega} + \frac{ret\,\epsilon^2\,\phi\,q1''[ti]}{\omega} - \right.$$
$$\frac{2\,ret^2\,\epsilon^2\,\phi\,q1''[ti]}{\omega} - \frac{3\,\epsilon^2\,q1'[ti]^2\,q1''[ti]}{2\,\omega^3} + \frac{4\,kin\,\epsilon^2\,\phi\,q2''[ti]}{\omega} -$$
$$\left.\frac{ret\,\epsilon^2\,\phi\,q2''[ti]}{\omega} + \frac{2\,ret^2\,\epsilon^2\,\phi\,q2''[ti]}{\omega}\right) == 0,$$
$$-\epsilon\,\phi\,\omega + 2\,\epsilon^2\,\phi\,\omega\,q1[ti] + 2\,\epsilon^2\,\phi\,\omega\,q2[ti] - \frac{q2''[ti]}{\omega} +$$
$$\Omega^2\left(\frac{4\,kin\,\epsilon^2\,\phi\,q1''[ti]}{\omega} - \frac{ret\,\epsilon^2\,\phi\,q1''[ti]}{\omega} + \frac{2\,ret^2\,\epsilon^2\,\phi\,q1''[ti]}{\omega} - \frac{3\,kin\,\epsilon^2\,\phi\,q2''[ti]}{\omega} + \right.$$
$$\left.\left.\frac{ret\,\epsilon^2\,\phi\,q2''[ti]}{\omega} - \frac{2\,ret^2\,\epsilon^2\,\phi\,q2''[ti]}{\omega} - \frac{3\,\epsilon^2\,q2'[ti]^2\,q2''[ti]}{2\,\omega^3}\right) == 0\right\}$$

we solve the non-relativistic equations and then find the correction

```
In[ ]:= eqsNonrel =
         Collect[eqs // Series[#, {Ω, 0, 0}] & // Normal, Ω] /. {q1 → q1nr, q2 → q2nr}
        (*the non-relativistic equations*)
        solNonrel =
          DSolve[{eqsNonrel[[1]], eqsNonrel[[2]], q1nr[0] ⩵ y1p, q2nr[0] ⩵ y2p, q1nr[t] ⩵ y1,
             q2nr[t] ⩵ y2}, {q1nr[ti], q2nr[ti]}, {ti}] /.
           ti → tii // Simplify; (*their solution*)
        g1[ti_] := (q1nr[tii] /. solNonrel[[1]] /. tii → ti) + Ω c1[ti];
        (*the solution for the relativistic equations to the relevant
          order would be the non-relativistic solution plus a correction*)
        g2[ti_] := (q2nr[tii] /. solNonrel[[1]] /. tii → ti) + Ω c2[ti];
        eqs2 = eqs /. {q1 → g1, q2 → g2} // mySeries // Refine // Simplify
        (*equation for the corrections*)
        solCor = DSolve[{eqs2[[1]], eqs2[[2]], c1[0] ⩵ 0, c2[0] ⩵ 0, c1[t] ⩵ 0, c2[t] ⩵ 0},
           {c1[ti], c2[ti]}, {ti}] (*their solution*)
```

$$Out[\ ]= \left\{-\epsilon\, \phi\, \omega + 2\, \epsilon^2\, \phi\, \omega\, \text{q1nr}[ti] + 2\, \epsilon^2\, \phi\, \omega\, \text{q2nr}[ti] - \frac{\text{q1nr}''[ti]}{\omega} \equiv 0,\right.$$
$$\left.-\epsilon\, \phi\, \omega + 2\, \epsilon^2\, \phi\, \omega\, \text{q1nr}[ti] + 2\, \epsilon^2\, \phi\, \omega\, \text{q2nr}[ti] - \frac{\text{q2nr}''[ti]}{\omega} \equiv 0\right\}$$

$$Out[\ ]= \left\{2\, \epsilon^2\, \phi\, \omega^2\, (\text{c1}[ti] + \text{c2}[ti]) \equiv \text{c1}''[ti],\ 2\, \epsilon^2\, \phi\, \omega^2\, (\text{c1}[ti] + \text{c2}[ti]) \equiv \text{c2}''[ti]\right\}$$

$$Out[\ ]= \{\{\text{c1}[ti] \to 0,\ \text{c2}[ti] \to 0\}\}$$

turns out that at this order there is no correction to the classical solutions.

```
In[ ]:= solCor = solCor /. ti → tii // Simplify;
        x1tii = g1[tii] /. solCor[[1]] // Simplify;
        x2tii = g2[tii] /. solCor[[1]] // Simplify;
        x1[ti_] := x1tii /. tii → ti ;
        x2[ti_] := x2tii /. tii → ti ;
```

Sanity check: initial and final locations are actually the y's, the trajectories actually solve the equations:

```
In[ ]:= eqs /. {q1 → x1, q2 → x2} // mySeries // Simplify
       x1[0] // mySeries // Simplify
       x2[0] // mySeries // Simplify
       x1[t] // mySeries // Simplify
       x2[t] // mySeries // Simplify
```

*Out[ ]=*
{True, True}

*Out[ ]=*
y1p

*Out[ ]=*
y2p

*Out[ ]=*
y1

*Out[ ]=*
y2

---

# Wavefunction

We would now like to compute the wavefunction $\tilde{\psi}^{fi}(y1, y2)$ as given in Eq. (9). We start with achieving the expression for the on-shell ( = stationary-phase approximated) action $\tilde{S}_a + \tilde{S}_G$, here Shb.

*In[ ]:=* ```
Lhb =  LGhb + Lunc /. {q1 → x1, q2 → x2} // mySeries // Simplify;
Shb1 = Integrate[Lhb, {ti, 0, t}] ;
Shb = Shb1 - (Shb1 /. {y1 → 0, y2 → 0, y1p → 0, y2p → 0}) // Expand // Refine //
   Simplify  // Collect[#, {y1, y2, y1p, y2p}] &
```

*Out[ ]=*

$$-\frac{1}{2} t\, y2p\, \epsilon\, \phi\, \omega + \frac{y1^4\, \epsilon^2\, \Omega^2}{8\, t^3\, \omega^3} - \frac{y1^3\, y1p\, \epsilon^2\, \Omega^2}{2\, t^3\, \omega^3} + \frac{y1p^4\, \epsilon^2\, \Omega^2}{8\, t^3\, \omega^3} + \frac{y2^4\, \epsilon^2\, \Omega^2}{8\, t^3\, \omega^3} - \frac{y2^3\, y2p\, \epsilon^2\, \Omega^2}{2\, t^3\, \omega^3} +$$

$$\frac{y2p^4\, \epsilon^2\, \Omega^2}{8\, t^3\, \omega^3} + y1p^2 \left(\frac{1}{3}\, t\, \epsilon^2\, \phi\, \omega + \frac{1 + 3\, \text{kin}\, \epsilon^2\, \phi\, \Omega^2 - \text{ret}\, \epsilon^2\, \phi\, \Omega^2 + 2\, \text{ret}^2\, \epsilon^2\, \phi\, \Omega^2}{2\, t\, \omega}\right) +$$

$$y2p^2 \left(\frac{1}{3}\, t\, \epsilon^2\, \phi\, \omega + \frac{1 + 3\, \text{kin}\, \epsilon^2\, \phi\, \Omega^2 - \text{ret}\, \epsilon^2\, \phi\, \Omega^2 + 2\, \text{ret}^2\, \epsilon^2\, \phi\, \Omega^2}{2\, t\, \omega}\right) +$$

$$y1^2 \left(\frac{1}{3}\, t\, \epsilon^2\, \phi\, \omega + \frac{3\, y1p^2\, \epsilon^2\, \Omega^2}{4\, t^3\, \omega^3} + \frac{1 + 3\, \text{kin}\, \epsilon^2\, \phi\, \Omega^2 - \text{ret}\, \epsilon^2\, \phi\, \Omega^2 + 2\, \text{ret}^2\, \epsilon^2\, \phi\, \Omega^2}{2\, t\, \omega}\right) +$$

$$y2^2 \left(\frac{1}{3}\, t\, \epsilon^2\, \phi\, \omega + \frac{3\, y2p^2\, \epsilon^2\, \Omega^2}{4\, t^3\, \omega^3} + \frac{1 + 3\, \text{kin}\, \epsilon^2\, \phi\, \Omega^2 - \text{ret}\, \epsilon^2\, \phi\, \Omega^2 + 2\, \text{ret}^2\, \epsilon^2\, \phi\, \Omega^2}{2\, t\, \omega}\right) +$$

$$y1p \left(-\frac{1}{2}\, t\, \epsilon\, \phi\, \omega + y2p \left(\frac{2}{3}\, t\, \epsilon^2\, \phi\, \omega + \frac{(-4\, \text{kin} - \text{ret}\, (-1 + 2\, \text{ret}))\, \epsilon^2\, \phi\, \Omega^2}{t\, \omega}\right)\right) +$$

$$y1 \left(-\frac{1}{2}\, t\, \epsilon\, \phi\, \omega - \frac{y1p^3\, \epsilon^2\, \Omega^2}{2\, t^3\, \omega^3} + y2 \left(\frac{2}{3}\, t\, \epsilon^2\, \phi\, \omega + \frac{(-4\, \text{kin} - \text{ret}\, (-1 + 2\, \text{ret}))\, \epsilon^2\, \phi\, \Omega^2}{t\, \omega}\right) + \right.$$

$$\left. y2p \left(\frac{1}{3}\, t\, \epsilon^2\, \phi\, \omega + \frac{(4\, \text{kin} + \text{ret}\, (-1 + 2\, \text{ret}))\, \epsilon^2\, \phi\, \Omega^2}{t\, \omega}\right) + \right.$$

$$\left. y1p \left(\frac{1}{3}\, t\, \epsilon^2\, \phi\, \omega - \frac{1 + 3\, \text{kin}\, \epsilon^2\, \phi\, \Omega^2 - \text{ret}\, \epsilon^2\, \phi\, \Omega^2 + 2\, \text{ret}^2\, \epsilon^2\, \phi\, \Omega^2}{t\, \omega}\right)\right) +$$

$$y2 \left(-\frac{1}{2}\, t\, \epsilon\, \phi\, \omega - \frac{y2p^3\, \epsilon^2\, \Omega^2}{2\, t^3\, \omega^3} + y1p \left(\frac{1}{3}\, t\, \epsilon^2\, \phi\, \omega + \frac{(4\, \text{kin} + \text{ret}\, (-1 + 2\, \text{ret}))\, \epsilon^2\, \phi\, \Omega^2}{t\, \omega}\right) + \right.$$

$$\left. y2p \left(\frac{1}{3}\, t\, \epsilon^2\, \phi\, \omega - \frac{1 + 3\, \text{kin}\, \epsilon^2\, \phi\, \Omega^2 - \text{ret}\, \epsilon^2\, \phi\, \Omega^2 + 2\, \text{ret}^2\, \epsilon^2\, \phi\, \Omega^2}{t\, \omega}\right)\right)$$

We now solve the y1p and y2p integrals while also performing the approximations. We denote for any function f the parts fExp and fPre such that $f(x) = fPre(x) \times e^{fExp(x)}$.

The initial state wfIn is a gaussian.

```
vars = {y1, y2, y1p, y2p};
wfIn = Exp[ - (y1p) ^ 2 / 2 - (y2p) ^ 2 / 2];

wf4vars = wfIn Exp[i Shb];(*the integrand in Eq. (9).*)
wf4vars = wf4vars // mySeries;
wf4varsExp = Exponent[wf4vars, ϵ];
wf4varsPre = wf4vars Exp[-wf4varsExp] ;
wf = myGauss[ wf4varsPre , wf4varsExp, {y1p, y2p}] // Simplify;
(*This is ψ~ fi(y1, y2)*)
wfExp = Exponent[wf, ϵ] // Simplify
wfPre = wf Exp[-wfExp]
wfExpC = wfExp // myConj // Simplify;(*complex conjugate*)
wfPreC = wfPre // myConj // Simplify;
```

*Out[ ]=*

$$\frac{i\,(y1^2 + y2^2)}{-2\,i + 2\,t\,\omega}$$

*Out[ ]=*

$$-\frac{1}{12\,(-i + t\,\omega)^5}$$

$\pi\,\bigl(-t^5\,\bigl(24 - 60\,y2\,\epsilon\,\phi + 6\,\epsilon^2\,\phi\,(8 + 3\,\phi) + y1^2\,\epsilon^2\,\phi\,(40 + 39\,\phi) + y2^2\,\epsilon^2\,\phi\,(40 + 39\,\phi) +$
$\quad 2\,y1\,\epsilon\,\phi\,(-30 + y2\,\epsilon\,(40 + 39\,\phi))\bigr)\,\omega^5 - 2\,i\,t^6\,\epsilon\,\phi$
$\quad \bigl(-6\,y2 + y1^2\,\epsilon\,(4 + 9\,\phi) + y2^2\,\epsilon\,(4 + 9\,\phi) + \epsilon\,(8 + 9\,\phi) + 2\,y1\,(-3 + y2\,\epsilon\,(4 + 9\,\phi))\bigr)\,\omega^6 +$
$\quad 3\,t^7\,\bigl(2 + y1^2 + 2\,y1\,y2 + y2^2\bigr)\,\epsilon^2\,\phi^2\,\omega^7 + 18\,i\,\epsilon^2\,\Omega^2 -$
$\quad 6\,t\,\omega\,\bigl(4 + \epsilon^2\,\bigl(6 + 3\,y1^2 + 3\,y2^2 - 12\,\text{kin}\,\phi + 4\,\text{ret}\,\phi - 8\,\text{ret}^2\,\phi\bigr)\,\Omega^2\bigr) -$
$\quad 3\,i\,t^2\,\omega^2\,\bigl(32 - 8\,y2\,\epsilon\,\phi + 6\,\epsilon^2\,\Omega^2 + y1^4\,\epsilon^2\,\Omega^2 + y2^4\,\epsilon^2\,\Omega^2 - 72\,\text{kin}\,\epsilon^2\,\phi\,\Omega^2 + 24\,\text{ret}\,\epsilon^2\,\phi\,\Omega^2 -$
$\quad\quad 48\,\text{ret}^2\,\epsilon^2\,\phi\,\Omega^2 + 2\,y1^2\,\epsilon^2\,\bigl(3\,\Omega^2 + \phi\,\bigl(4 - 6\,\text{kin}\,\Omega^2 + 2\,\text{ret}\,\Omega^2 - 4\,\text{ret}^2\,\Omega^2\bigr)\bigr) +$
$\quad\quad 2\,y2^2\,\epsilon^2\,\bigl(3\,\Omega^2 + \phi\,\bigl(4 - 6\,\text{kin}\,\Omega^2 + 2\,\text{ret}\,\Omega^2 - 4\,\text{ret}^2\,\Omega^2\bigr)\bigr) +$
$\quad\quad 8\,y1\,\epsilon\,\phi\,\bigl(-1 + y2\,\epsilon\,\bigl(2 + 4\,\text{kin}\,\Omega^2 - \text{ret}\,\Omega^2 + 2\,\text{ret}^2\,\Omega^2\bigr)\bigr)\bigr) +$
$\quad 4\,t^3\,\omega^3\,\bigl(36 - 21\,y2\,\epsilon\,\phi + 4\,\epsilon^2\,\phi - 54\,\text{kin}\,\epsilon^2\,\phi\,\Omega^2 + 18\,\text{ret}\,\epsilon^2\,\phi\,\Omega^2 -$
$\quad\quad 36\,\text{ret}^2\,\epsilon^2\,\phi\,\Omega^2 + 3\,y1^2\,\epsilon^2\,\phi\,\bigl(6 + \phi - 6\,\text{kin}\,\Omega^2 + 2\,\text{ret}\,\Omega^2 - 4\,\text{ret}^2\,\Omega^2\bigr) +$
$\quad\quad 3\,y2^2\,\epsilon^2\,\phi\,\bigl(6 + \phi - 6\,\text{kin}\,\Omega^2 + 2\,\text{ret}\,\Omega^2 - 4\,\text{ret}^2\,\Omega^2\bigr) +$
$\quad\quad 3\,y1\,\epsilon\,\phi\,\bigl(-7 + 2\,y2\,\epsilon\,\bigl(6 + \phi + 8\,\text{kin}\,\Omega^2 - 2\,\text{ret}\,\Omega^2 + 4\,\text{ret}^2\,\Omega^2\bigr)\bigr)\bigr) +$
$\quad 2\,i\,t^4\,\omega^4\,\bigl(48 - 54\,y2\,\epsilon\,\phi + 3\,\epsilon^2\,\phi\,\bigl(8 + \phi - 12\,\text{kin}\,\Omega^2 + 4\,\text{ret}\,\Omega^2 - 8\,\text{ret}^2\,\Omega^2\bigr) +$
$\quad\quad 2\,y1^2\,\epsilon^2\,\phi\,\bigl(20 + 9\,\phi - 9\,\text{kin}\,\Omega^2 + 3\,\text{ret}\,\Omega^2 - 6\,\text{ret}^2\,\Omega^2\bigr) +$
$\quad\quad 2\,y2^2\,\epsilon^2\,\phi\,\bigl(20 + 9\,\phi - 9\,\text{kin}\,\Omega^2 + 3\,\text{ret}\,\Omega^2 - 6\,\text{ret}^2\,\Omega^2\bigr) +$
$\quad\quad 2\,y1\,\epsilon\,\phi\,\bigl(-27 + 2\,y2\,\epsilon\,\bigl(20 + 9\,\phi + 12\,\text{kin}\,\Omega^2 - 3\,\text{ret}\,\Omega^2 + 6\,\text{ret}^2\,\Omega^2\bigr)\bigr)\bigr)\bigr)$

calculate the norm squared of the state:

*In[ ]:=*
```
preee = wfPreC wfPre // mySeriesSq // Simplify;
normsq = myGauss[preee, wfExpC + wfExp, {y1, y2}];
normsq = normsq // Simplify;
```

# Generalized Entanglement Measure

we now calculate the entanglement measure in Eq. (28) after replacing the norm square in the formula by multiplication with the conjugate:

$$\left|\int \psi(y1, y2)\, \psi^*(y'1, y2)\, dy2\right|^2 = \int \psi^*(y1, y2)\, \psi(y'1, y2)\, \psi(y1, y'2)\, \psi^*(y'1, y'2)\, dy2\, dy'2$$

*In[ ]:=*
```
(*each of the four multipicants is slightly
   different in where is the ' and whether it is conjugate*)
wfPre1p = wfPre /. y1 → y1p;
wfExp1p = wfExp /. y1 → y1p;
wfPre2p = wfPre /. y2 → y2p;
wfExp2p = wfExp /. y2 → y2p;
wfPreC12p = wfPreC /. {y1 → y1p, y2 → y2p};
wfExpC12p = wfExpC /. {y1 → y1p, y2 → y2p};

(*again, preform the integral after getting
   to the form myGauss accepts: f(x)=fPre(x)*e^fExp(x)*)
thisPre = wfPre1p * wfPreC * wfPreC12p * wfPre2p // mySeriesSq;
Print["pre series"];
thisExp = wfExp1p + wfExpC + wfExpC12p + wfExp2p // mySeriesSq // FullSimplify;
Print["exp simp"];
int = myGauss[thisPre, thisExp, {y1, y2, y1p, y2p}];
Print["int"];
int = int // Simplify ;
Print["int simp"];
int = int // mySeriesSq;
Print["int ser"];
pre series
exp simp
int
int simp
int ser
```

finally, the generalized entanglement measure from Eq. (28) is calculated to obtain:

*In[ ]:=*
```
gem = 2 (1 - int / normsq^2 ) // Sqrt // mySeries // FullSimplify;
gem = gem // Collect[#, {Ω, kin}] &
```

*Out[ ]=*

$$\frac{2}{3} t \,\epsilon^2\, \phi\, \omega \sqrt{9 + 3\, t^2\, \omega^2 + t^4\, \omega^4} + \left( -\frac{4\, kin\, t\, \epsilon^2\, \phi\, \omega\, (-3 + t^2\, \omega^2)}{\sqrt{9 + 3\, t^2\, \omega^2 + t^4\, \omega^4}} - \frac{ret\, (-1 + 2\, ret)\, t\, \epsilon^2\, \phi\, \omega\, (-3 + t^2\, \omega^2)}{\sqrt{9 + 3\, t^2\, \omega^2 + t^4\, \omega^4}} \right) \Omega^2$$

the different contributions, see Supplementary Material:

*In[ ]:=* **gem /. {kin → 1, ret → 0}**
   (*with correction due to gravitation of kinetic energy only*)

*Out[ ]=*
$$\frac{2}{3} t \epsilon^2 \phi \omega \sqrt{9 + 3 t^2 \omega^2 + t^4 \omega^4} - \frac{4 t \epsilon^2 \phi \omega \left(-3 + t^2 \omega^2\right) \Omega^2}{\sqrt{9 + 3 t^2 \omega^2 + t^4 \omega^4}}$$

*In[ ]:=* **gem /. {kin → 0, ret → 1}** (*with correction due to retardation only*)

*Out[ ]=*
$$\frac{2}{3} t \epsilon^2 \phi \omega \sqrt{9 + 3 t^2 \omega^2 + t^4 \omega^4} - \frac{t \epsilon^2 \phi \omega \left(-3 + t^2 \omega^2\right) \Omega^2}{\sqrt{9 + 3 t^2 \omega^2 + t^4 \omega^4}}$$

*In[ ]:=* **gem /. {kin → 1, ret → 1}**
   (*with both corrections: the first GR correction to the entanglement*)

*Out[ ]=*
$$\frac{2}{3} t \epsilon^2 \phi \omega \sqrt{9 + 3 t^2 \omega^2 + t^4 \omega^4} - \frac{5 t \epsilon^2 \phi \omega \left(-3 + t^2 \omega^2\right) \Omega^2}{\sqrt{9 + 3 t^2 \omega^2 + t^4 \omega^4}}$$



% \includepdf[pages={1,{},2-}]{paths.pdf}
% \includepdf[pages={1,{},2-}]{rel.pdf}

\end{document}